\documentclass[12pt,draftclsnofoot,peerreviewca,letterpaper,onecolumn]{IEEEtran}

\usepackage{cite}
\usepackage{graphicx,color,epsfig,rotating}
\usepackage{amsfonts,amsmath,amssymb,bbm}
\usepackage{algorithm, algorithmic}
\usepackage{subfigure}
\usepackage{amsmath}
\usepackage{cite}
\usepackage{mdwtab}
\usepackage{subfigure}
\usepackage{placeins}
\usepackage{psfrag, graphicx}
\usepackage[latin1]{inputenc}
\usepackage{amssymb}
\usepackage{makeidx}
\usepackage{epstopdf}
\usepackage{scalerel}
\usepackage{comment}
\newfont{\bbb}{msbm10 scaled 700}

\newfont{\bb}{msbm10 scaled 1100}

\newcommand{\Mc}{{\cal M}}

\newcommand{\eqdef}{\stackrel{\Delta}{=}}

\newcommand{\be}{\begin{equation}}
\newcommand{\ee}{\end{equation}}
\newcommand{\bea}{\begin{eqnarray}}
\newcommand{\eea}{\end{eqnarray}}

\newtheorem{defn}{Definition}
\newtheorem{theorem}{Theorem}
\newtheorem{remark}{Remark}

\begin{document}
\setcounter{page}{1}
\title{A New Combinatorial Coded Design for Heterogeneous Distributed Computing\thanks{This manuscript was partially presented in the conference papers \cite{woolsey2020icc,woolsey2019coded}.}}

\author{Nicholas Woolsey,~\IEEEmembership{Student Member,~IEEE}, Rong-Rong Chen,~\IEEEmembership{Member,~IEEE},\\ and Mingyue Ji,~\IEEEmembership{Member,~IEEE}
\thanks{The authors are with the Department of Electrical Engineering,
University of Utah, Salt Lake City, UT 84112, USA. (e-mail: nicholas.woolsey@utah.edu, rchen@ece.utah.edu and mingyue.ji@utah.edu)}
}

\if{0}
\author{
    \IEEEauthorblockN{ Nicholas Woolsey,
		Rong-Rong Chen, and Mingyue Ji }
	\IEEEauthorblockA{Department of Electrical and Computer Engineering, University of Utah\\
		Salt Lake City, UT, USA\\
		Email: \{nicholas.woolsey@utah.edu,
		 rchen@ece.utah.edu,
		mingyue.ji@utah.edu\}}
\thanks{The authors are with the Department of Electrical Engineering,
University of Utah, Salt Lake City, UT 84112, USA. (e-mail: nicholas.woolsey@utah.edu, rchen@ece.utah.edu and mingyue.ji@utah.edu)}
}
\fi

\maketitle

\vspace{-0.5cm}

\begin{abstract}
Coded Distributed Computing (CDC) introduced by Li {\em et al.} in 2015 offers an efficient approach to trade computing power to reduce the communication load in general distributed computing frameworks such as MapReduce and Spark. In particular, increasing the computation load in the Map phase by a factor of $r$ can create coded multicasting opportunities to reduce the communication load in the Shuffle phase by the same factor. However, the CDC scheme is designed for the homogeneous settings, where the storage, computation load and communication load on the computing nodes are the same. In addition, it requires an exponentially large number of input files (data batches), reduce functions and multicasting groups relative to the number of nodes to achieve the promised gain. 
We address the CDC limitations by proposing a novel CDC approach based on a combinatorial design, which accommodates heterogeneous networks where nodes have varying storage and computing capabilities. In addition, the proposed approach requires an exponentially less number of input files 
 compared to the original CDC scheme proposed by Li {\em et al.} Meanwhile, the resulting computation-communication trade-off maintains the multiplicative gain compared to conventional uncoded unicast and asymptotically achieves the optimal performance proposed by Li {\em et al.} 
\end{abstract}

\begin{IEEEkeywords}
Coded Distributed Computing, Communication load, Computation load, Coded multicasting, Heterogeneity, Low-Complexity  
\end{IEEEkeywords}



\section{Introduction}
\label{section: intro}
In recent years, coding has been reinvented for solving problems in distributed computing systems from different perspectives such as straggler mitigation \cite{lee2017speeding,dutta2016short,ferdinand2016anytime,yu2018straggler,bitar2017minimizing,karakus2017straggler,halbawi2017improving,suh2017matrix,mallick2018rateless,maity2018robust,Aktas2018straggler,wang2018fundamental,ye2018communication,wan2020distributed}, 
data shuffling \cite{lee2017speeding,attia2019shuffling,Elmahdy2020shuffling,wan2020shuffling}, and robustness \cite{chen2018draco}.
In particular, Coded Distributed Computing (CDC), introduced in \cite{li2018fundamental}, offers an efficient approach to reduce the communication load in distributed computing networks such as MapReduce \cite{dean2008mapreduce}. 
In this type of distributed computing network, in order to compute the output functions, the computation is decomposed into ``Map" and ``Reduce" phases. First, each computing node computes intermediate values (IVs) using  local input data files according to the designed Map functions. Then, computed IVs are exchanged among  computing nodes 
and nodes use these IVs as input to the designed reduce functions to compute output functions.
The operation of exchanging IVs is called ``data shuffling" and occurs during the ``Shuffle" phase. This severely limits the performance of distributed computing applications due to the very high transmitted traffic load \cite{li2018fundamental}.

In \cite{li2018fundamental}, by formulating and characterizing a fundamental tradeoff between ``computation load" in the Map phase and ``communication load" in the Shuffle phase, Li  {\em et al.} demonstrated that these two quantities are inversely proportional to each other. This means that if each intermediate value is computed at $r$ carefully chosen nodes,
then the communication load in the Shuffle phase can be reduced by a factor of $r$. CDC achieves this multiplicative gain in the Shuffle phase by leveraging
coding opportunities created in the Map phase by strategically placing the input files among the computing nodes.
However, there are a few limitations of the CDC scheme in \cite{li2018fundamental}. First, it requires an exponentially large number of input files and reduce functions relative to the number of computing nodes. 
In some cases, the number of files and functions becomes unrealistic and the promised again cannot be achieved in practice. Second, there is an exponential number of multicasting groups compared to the number of nodes and the computation load. When implementing CDC in \cite{li2018fundamental}, the execution time of the code generation step is proportional to the number of multicasting groups. This counteracts the benefits of CDC in reducing overall execution time. Third, the CDC scheme assumes the computing network is homogeneous in that each computing node has the same computation and storage resources which limits its effectiveness on heterogeneous computing networks.

Some other aspects of CDC have been investigated in the literature. In \cite{ezzeldin2017communication}, Ezzeldin {\em et al.} revisited the computation-communication tradeoff by computing only necessary IVs in each node. The authors proposed a lower bound on the corresponding computation load via a heuristic scheme, which achieves the lower bound under certain parameter regimes.  In \cite{song2017benefit}, Song {\em et al.} considered the case where each computing node has access to a random subset of  input files and the system is asymmetric. This means that not all output functions depend on the entire data set and we can decide which node computes which functions. The corresponding communication load was characterized. Later, in \cite{prakash2018coded}, Prakash {\em et al.} extended CDC to graph analytics of Erd\"os-R\'enyi graphs, where the computation at each vertex uses data
only from the adjacent vertices. 
In \cite{Srinivasavaradhan2018distributed}, Srinivasavaradhan {\em et al.} considered the CDC design under a random network topology following a Erd\"os-R\'enyi random graph model.
In \cite{konstantinidis2018leveraging}, the Konstantinidis {\em et al.} used resolvable designs to reduce the necessary number of files, functions, and number of multicasting groups. Furthermore, they implemented  new designs to demonstrate an overall reduction in execution time compared to  implementations of \cite{li2018fundamental} for some cases.

Thus far, all aforementioned prior works have assumed the CDC network to be homogeneous, that is, the computing nodes of the network have the same amount of storage, computation, and communication resources. Understanding the performance potential and finding achievable designs for heterogeneous networks remains an open problem.
The authors in \cite{kiamari2017Globecom} derived a lower bound for the communication load for a CDC network where nodes have varying storage or computing capabilities. The proposed achievable scheme achieves the information-theoretical optimality of the minimum communication load for a system of $3$ nodes. The authors also demonstrated that the parameters of a heterogeneous CDC network can be translated into an optimization problem to find an efficient Map and Shuffle phase design.
In \cite{shakya2018distributed}, the authors studied CDC networks with $2$ and $3$ computing nodes where nodes have varying communication load constraints to find a lower bound on the minimum computation load. These works mainly focus on the heterogeneous placement of the files in the Map phase, however, nodes are assumed to have a homogeneous reduce function assignment. The authors of \cite{xu2019heterogeneous} explore the concept of semi-random file placement and function assignment and develop a heterogeneous computing scheme which can operate on a computing network with arbitrary heterogeneous storage and computation requirements. However, the number of necessary files and functions of this scheme are unclear as files and functions are assigned as fractions of the entire file library and function set, respectively.

Our contributions in this paper are as follows. 
\begin{itemize}
\begin{comment}
\item First, we  propose a novel CDC 
 approach based on a combinatorial design, called {\em hypercube}, for the homogeneous networks, which 
requires an exponentially less number of input files 
and multicasting groups as compared to that in \cite{li2018fundamental}. 
Meanwhile, the resulting computation-communication trade-off maintains the multiplicative gain compared to conventional uncoded MapReduce and achieves the optimal trade-off proposed in \cite{li2018fundamental} asymptotically. 
\end{comment}
\item First, we establish a novel combinatorial framework for CDC that exploits elegant geometric structures--  {\em hypercube} for  homogeneous networks and {\em hypercuboid} for heterogeneous networks, to optimize the tradeoff of communication and computing for such networks. The proposed designs require an exponentially less number of input files and multicasting groups as compared to that in \cite{li2018fundamental}. Meanwhile, the resulting computation-communication trade-off maintains the multiplicative gain compared to conventional uncoded MapReduce and achieves the optimal trade-off proposed in \cite{li2018fundamental} asymptotically.
\item Second, 
the proposed hypercuboid design can accommodate
large heterogeneous 
CDC networks where nodes have 
varying storage and computing capabilities. 
This is achieved by the combinatorial design of a heterogeneous network (hypercuboid) consisting of multiple interleaved homogeneous networks (hypercubes) with varying dimensions and the design of efficient file mapping and data shuffle schemes across them. 
Another novelty of the proposed design is to assign more output functions to  nodes with more storage space and computing resources. This is in contrast to  previous work where each node is assigned by the same number of output functions \cite{kiamari2017Globecom}. 
Based on the proposed file and function assignments, we characterize an information theoretic converse bound, which is tight within a constant factor. According to our knowledge, this is the first work that develops an explicit and systematic heterogeneous CDC design with optimality guarantees 
under certain network parameters. 

\item Third, 
this work shows 
that network heterogeneity can actually reduce the communication load and thus, the fundamental tradeoff of \cite{li2018fundamental} no longer applies in this setting.\footnote{A similar phenomenon was also observed in \cite{xu2019heterogeneous}.} 
For large heterogeneous networks, we show that the proposed heterogeneous design can achieve a communication load that is strictly less than that of an equivalent homogeneous network.  of \cite{li2018fundamental}.
\end{itemize}

The remainder of this paper is outlined as follows. In Section \ref{sec: Network Model and Problem Formulation}, we present the network model and problem formulation. Then, we present 
the proposed combinatorial CDC design and discuss its performance in Section \ref{sec: Hypercube Caching Network Approach} for the homogeneous case and in Section \ref{sec: hets1} for the more general heterogeneous case. 
In Section \ref{sec: Discussion}, we compare our design to the state-of-the-art design of \cite{li2018fundamental}. Concluding remarks are provided in Section \ref{sec: Conclusion}.

\paragraph*{Notation Convention}
We use $|\cdot|$ to represent the cardinality of a set or the length of a vector. 
Also $[n] := \{1,2,\ldots,n\}$ for some $n\in\mathbb{Z}^+$, where $\mathbb{Z}^+$ is the set of all positive integers, and $\oplus$ represents bit-wise XOR. 

\section{Network Model and Problem Formulation}
\label{sec: Network Model and Problem Formulation}

The network model is adopted from \cite{li2018fundamental}. We consider a distributed computing network where a set of $K$ nodes, labeled as $[K]=\{1, \ldots , K \}$, have the goal of computing $Q$ output functions and computing each function requires access to all $N$ input files. The  input files, $\{w_1 , \ldots , w_N \}$, are assumed to be of equal size of $B$ bits each. The set of $Q$ output functions is denoted by $\{ \phi_1 , \ldots \phi_Q\}$. Each node $k\in [K] $ is assigned to compute a subset of output functions, denoted by $\mathcal{W}_k \subseteq [Q] $ ({\em function assignment}). The result of output function $i \in [Q]$ is $u_i = \phi_i \left( w_1, \ldots , w_N \right)$.

Alternatively, an output value of the targeted function $i$ can be computed  using the composition of ``Map" and ``Reduce" functions as follows. 
\be
u_i = h_i \left( g_{i,1}\left( w_1 \right), \ldots , g_{i,N}\left( w_N \right) \right),
\ee
 where for every output function $i$ there exists a set of $N$ Map functions $g_{i,j}(\cdot), i \in [Q], j \in [N]$ 
 and one Reduce function $h_i(\cdot), i \in [Q]$. Furthermore, we define the output of the Map function, $v_{i,j}=g_{i,j}\left( w_j \right)$, as the intermediate value (IV) resulting from performing the Map function for output function $i$ on file $w_j$. There are 
 $QN$ intermediate values in total and each is assumed to be size $T$ bits.

The MapReduce distributed computing framework allows nodes to compute output functions without having access to all $N$ files. Instead, each node $k$ has access to $M_k$ out of the $N$ files and we define the set of files available to node $k$ as $\mathcal{M}_k \subseteq \{ w_1, \ldots , w_N\}$ ({\em file mapping}). 
Collectively, the nodes use the Map functions to compute every IV in the {\em Map} phase at least once. Then, in the {\em Shuffle} phase, nodes multicast the computed IVs among one another via a shared link ({\em shuffle method}). 
The Shuffle phase is necessary so that each node can receive the necessary IVs that it could not compute itself. Finally, in the {\em Reduce} phase, nodes use the reduce functions with the appropriate IVs as inputs to compute the assigned output functions. 

Throughout this paper, we consider the following design options. First, we assume 
each computing node computes all possible IVs from locally available files. This means that $|\Mc_k|$ represents both storage space and computation load of each node. 
Second, we consider 
the design scenario such that
each of the $Q$ Reduce functions is 
computed exactly once ($s=1$) at one node
 and $|\mathcal{W}_i \cap \mathcal{W}_j| = 0$ for $i\neq j$, 
where $s$ is defined as the number of nodes which calculate each Reduce function.\footnote{The scenario of $s>1$, meaning that each of the $Q$ Reduce function is computed at multiple nodes, is called {\it cascaded} distributed computing, introduced in \cite{li2018fundamental}. In this paper, we do not consider this case.} 
Third,
we consider the general scenario where each computing node can have heterogeneous storage space and computing resource, or heterogenous size of $\Mc_k$ and $\mathcal{W}_k, \; \forall k \in [K]$. 
The proposed schemes accommodate heterogeneous networks in that nodes can be assigned a varying number of files and functions. 

This distributed computing network design yields two important performance parameters: the computation load, $r$, and the communication load, $L$. The computation load is defined as the number of times each IV is computed among all computing nodes, or $r = \frac{1}{N}\sum_{k=1}^K|\Mc_k|$. In other words, the computation load is the number of IVs computed in the Map phase normalized by the total number of unique IVs, $QN$. The communication load is defined as the amount of traffic load (in bits) among all the nodes in the Shuffle phase 
normalized by $QNT$. 
\begin{defn}
The optimal communication load is defined as 
\be
L^*(r) \eqdef \inf\{L: (r, L) \text{ is feasible}\}.
\ee
\end{defn}

\section{Homogeneous Hypercube Computing Approach}
\label{sec: Hypercube Caching Network Approach}


 In this section, we describe the proposed homogeneous CDC design based on the {\em hypercube} combinatorial structure. Our schemes are defined by 
  {\it node grouping}, {\it file mapping}, {\it function assignment} and {\it shuffle method}. Two detailed examples, one for two-dimensional, and one for three-dimensional, are provided to illustrate the fundamental principles of the proposed design. These will be extended to  the more general heterogeneous CDC scheme in Section \ref{sec: hets1}.

In this section, we consider the scenario where 
the network is homogeneous. In other words, each node is assigned the same number of files and reduce functions. Also, 
every reduce function is computed exactly once at one node ($s=1$). 
Every node computes a set of $\eta_2$ distinct functions and $Q=\eta_2 K$ where $\eta_2 \in \mathbb{Z}^+$.
The novel combinatorial hypercube design 
splits the nodes into $r$ disjoint sets each of size $\frac{K}{r}$ and batches of $\eta_1$ files are assigned to one node from each set.\footnote{This scheme can be classified as a resolvable design for CDC, which 
was introduced in \cite{konstantinidis2018leveraging}. In addition, it also falls into the general framework of  
the Placement Delivery Array (PDA) designed for Device-to-Device coded caching  \cite{wang2017placement}. 
} This is analogous to constructing a hypercube lattice of dimension $r$ with the length of each side $\frac{K}{r}$ to describe the file placement at the nodes. We use this hypercube approach to better illustrate the examples of our new combinatorial design.
We show that the required number of files is $N = \eta_1 \left( \frac{K}{r} \right)^r$ where $\eta_1 \in \mathbb{Z}^+$ and the number of multicasting groups is $G=\left( \frac{K}{r} \right)^r$. We first present a 2-dimension (a plane) example where $r=2$.

\subsection{2-Dimension Example}
\label{sec: 2D example}
\begin{figure}
\centering
\centering \includegraphics[width=9cm, height=6.3cm]{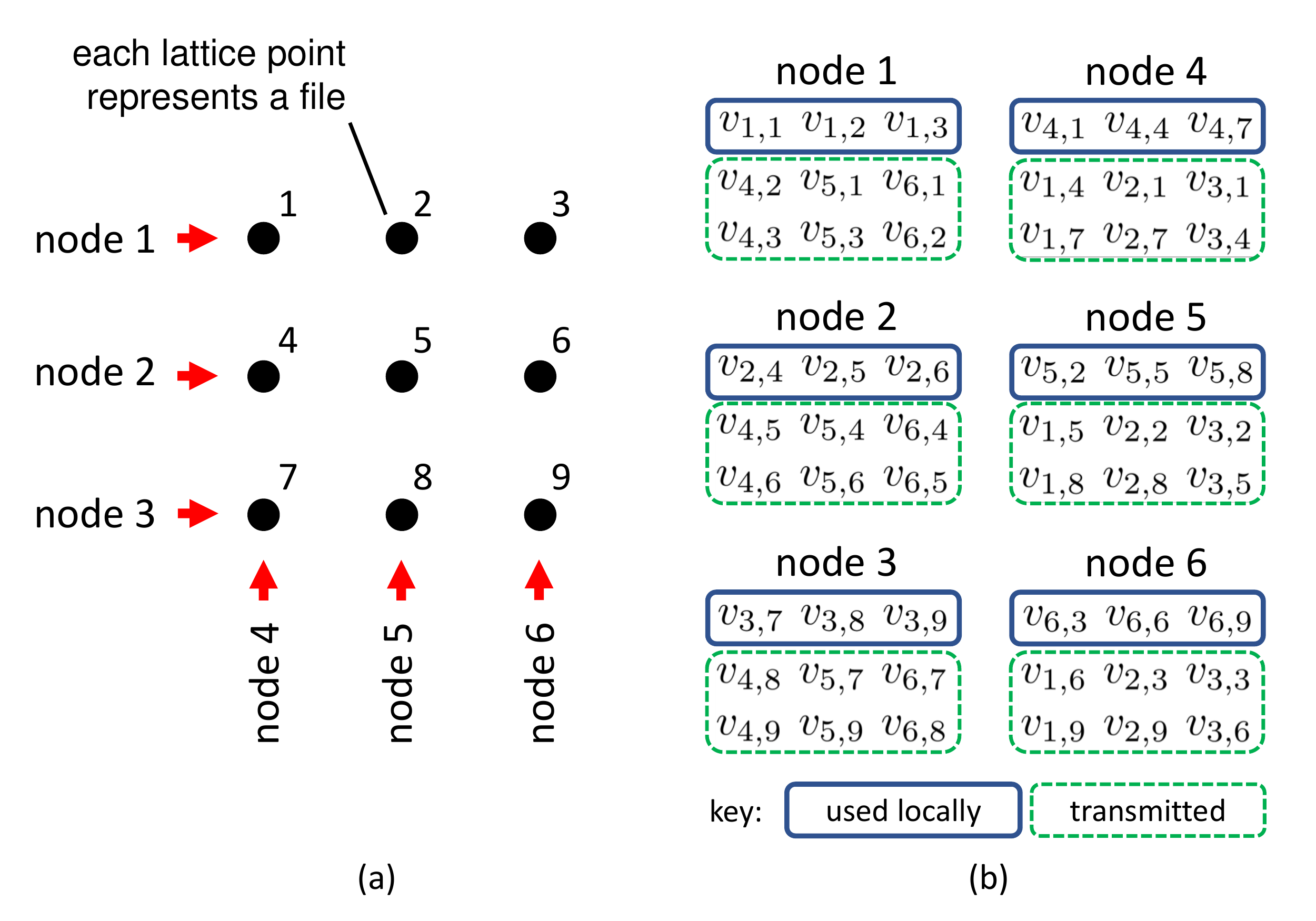}
\vspace{-0.7cm}
\caption{~\small (a) Lattice plane that defines file availability amongst the $K=6$ computing nodes. Each lattice point represents a file and each node has a set of files available to it represents by a horizontal or vertical line of lattice points. (b) The IVs used locally and transmitted by each node.}
\label{fig: 2d fig}
\vspace{-0.4cm}
\end{figure}

In this example, we propose a distributed computing network based on a $r=2$ dimensional hypercube (a plane) lattice where each side has length $\frac{K}{r}=3$. There are $K=6$ computing nodes each of which has access to $\frac{1}{3}$ of the file library. Each lattice point represents a file and each node has a set of files available to it represented by a line of lattice points as shown in Fig.~\ref{fig: 2d fig}(a). 
Specifically, there are two set of nodes: $\mathcal K_1=\{1,2,3\} $ and $\mathcal K_2=\{4,5,6\} $. Each node in $\mathcal K_1$ (or $\mathcal K_2$) has access to three files, represented by three lattice points along a horizontal (or vertical) line. For instance, node 1 in  $\mathcal K_1$ has access to three files $w_1$, $w_2$ and $w_3$ along the top horizontal line. Similarly, node 5  in $\mathcal K_2$ has access to three files  $w_2$, $w_5$ and $w_8$, along the middle vertical line. 
Each node is responsible for computing one out of the $Q=6$ reduce functions in the Reduce phase. More specifically, node $i$ computes reduce function $i$.

\if
In the Map phase, nodes compute intermediate values which are necessary to compute the reduce function they are responsible for. These intermediate values do not have to be transmitted and do not contribute to the communication load. For example, as shown in Fig.~\ref{fig: 2d fig}(b), node $1$ computes $v_{1,1}$, $v_{1,2}$ and $v_{1,3}$ and node 5 computes $v_{5,2}$, $v_{5,5}$ and $v_{5,8}$. 
Next, we consider all possible pairs of nodes, termed node groups, consisting of one node from $\mathcal K_1$ and one node from 
$\mathcal K_2$. 
For instance, node $1$ can form node groups with any of the three nodes in $\mathcal K_2=\{4,5,6\}$ and node $5$ can form node groups with  any of the three nodes in $\mathcal K_1=\{1,2,3\}$.
Within each node group of two nodes, each node  computes the intermediate values needed by the other node in the group. For the node group of  $\{1,5\}$, node $1$ computes $v_{5,1}$ and $v_{5,3}$ and transmits these values to node $5$.
Notice that, node $5$ is incapable of computing these intermediate values itself because it does not have access to files $w_1$ and $w_3$. Similarly, node $5$ computes $v_{1,5}$ and $v_{1,8}$ and transmits these intermediate values to node $1$ because node 1 does not have access to files $w_1$ and $w_8$. Fig.~\ref{fig: 2d fig}(b) shows the intermediate values computed at each node. For instance, node 1 computes $v_{1,1}$,  $v_{1,2}$, $v_{1,3}$ because node 1 is assigned to compute reduce function $1$ and it has access to files $w_1, w_2, w_3$. Node 1 can form three node groups with node 4, 5 or 6 in $\mathcal K_2$.  Thus, it will transmit intermediate values $v_{4,2}$ and $v_{4,3}$ to node 4, $v_{5,1}$ and $v_{5,3}$ to node 5, and $v_{6,1}$ and $v_{6,3}$ to node 6, respectively. 
On the other hand, node 1 will receive its requested intermediate values $v_{1,4}$ and $v_{1,7}$ from node 4,  $v_{1,5}$ and $v_{1,8}$ from node 5, and 
 $v_{1,6}$ and $v_{1,9}$ from node 6. This allows node 1 to obtain all the intermediate values necessary for computing reduction function $1$. In general, by considering all possible node groups, each node receives an intermediate value {\color{green}[] NW: Shall we use the acronym IV?]} for every file that it does not have. We can see this is true by recognizing that a node consecutively pairs with
the three nodes in either $\mathcal K_1$ or  $\mathcal K_2$, and the nodes in
either  $\mathcal K_1$ or  $\mathcal K_2$ collectively has access to all the files.
\fi

In the Map phase, nodes compute all $Q=6$ IVs from each locally available file. Some IVs are  necessary to compute the locally assigned reduce function. For example, as shown in Fig.~\ref{fig: 2d fig}(b), node $1$ computes $v_{1,1}$, $v_{1,2}$ and $v_{1,3}$ and node 5 computes $v_{5,2}$, $v_{5,5}$ and $v_{5,8}$. These IVs do not have to be transmitted and do not contribute to the communication load.
However, other IVs are transmitted between nodes. 
We consider all possible pairs of nodes, termed node groups, consisting of one node from $\mathcal K_1$ and one node from 
$\mathcal K_2$. 
For instance, nodes $1$ and $5$ form node groups with each of the three nodes in $\mathcal K_2=\{4,5,6\}$ or $\mathcal K_1=\{1,2,3\}$, respectively.
For the node group of  $\{1,5\}$, node $1$ has computed $v_{5,1}$ and $v_{5,3}$ and transmits these IVs to node $5$. 
Notice that, node $5$ is incapable of computing these IVs itself because it does not have access to files $w_1$ and $w_3$. Similarly, node $5$ has computed $v_{1,5}$ and $v_{1,8}$ and transmits these IVs to node $1$ because node 1 does not have access to files $w_1$ and $w_8$. Fig.~\ref{fig: 2d fig}(b) also shows the IVs transmitted by each node. For example, node $1$ will transmit IVs $v_{4,2}$ and $v_{4,3}$ to node $4$, $v_{5,1}$ and $v_{5,3}$ to node $5$ and $v_{6,1}$ and $v_{6,3}$ to node $6$. 
On the other hand, node 1 will receive its requested IVs $v_{1,4}$ and $v_{1,7}$ from node $4$,  $v_{1,5}$ and $v_{1,8}$ from node $5$, and 
 $v_{1,6}$ and $v_{1,9}$ from node $6$. Therefore, node $1$ obtains all the IVs necessary for computing reduce function $1$. In general, by considering all possible node groups, each node receives an IV for every file that it does not have. We can see this is true by recognizing that a node consecutively pairs with
the three nodes in either $\mathcal K_1$ or  $\mathcal K_2$, and the nodes in
either  $\mathcal K_1$ or  $\mathcal K_2$ collectively have access to all the files.

Throughout this paper, we mainly consider the case where each node computes all IVs from its available files similar to the original CDC work \cite{li2018fundamental}. In this example, each IV is computed twice and $r=2$ since each file is assigned to $2$ nodes. In general, the computation load $r$ is equivalent to the dimension of the hypercube which defines the file placement. {Note that, nodes will compute some IVs that are never used to transmit, decode or compute a reduce function.} From~\ref{fig: 2d fig}(b) shows the IVs computed by each node that are utilized. 
Each node computes $3$ IVs which are necessary for its own reduce function. Also, each node participates in $3$ node pairs for which it needs to compute $2$ IVs to transmit to the other node in the pair. {In some applications, it may be possible for nodes to compute a select set of IVs to reduce the computation load as presented in \cite{ezzeldin2017communication,woolsey2018new}.}

In this toy example, we only consider unicasting, therefore, the communication load is equivalent to the uncoded scenario and $L=\frac{2}{3}$, or the fraction of files not available at each node. This can be verified by recognizing that there are $9$ pairs of nodes for which $2$ IVs are transmitted from each node for each pair. In total, $36$ of the $54$ IVs are transmitted and $L=\frac{2}{3}$. In later examples, we will show how this 
scheme can be expanded to utilize coded multicasting and outperform the uncoded CDC scheme. 

\begin{remark}
Interestingly, although the general scheme generalized from this example is equivalent to the unicast in this case, we observe that there actually exist multicasting opportunities in this example. For instance, node $1$ could transmit $v_{4,2}\oplus v_{5,1}$ to nodes $4$ and $5$ (assuming that node $4$ and $5$ compute $v_{5,1}$ and $v_{4,2}$, respectively). In fact, all IVs could be transmitted in coded pairs where a node along one dimension transmits to $2$ nodes aligned along the other dimensions, which would reduce the communication load by half.\footnote{This is similar to the scheme outlined in \cite{wang2017placement,9133151} for the analogous coded caching problem. However, as we will see for other examples and as discussed in \cite{wang2017placement}, this scheme does not achieve a multiplicative gain for $r>2$.} 
\end{remark}

In the following, we describe the general scheme for the proposed combinatorial design which expands for the case when $r>2$.

\subsection{General Homogeneous Scheme} 
\label{sec: genscheme_s1}

In this subsection, we will introduce the general homogeneous scheme for $s=1$ step by step as follows. 


{\bf Node Grouping 1}: Let $\mathcal K=\{ 1, 2, \cdots, K \}$ denote the set of $K$ nodes. Assume that $\mathcal K$ is split into $r$ equal-sized  disjoint sets
$\mathcal{K}_1,\ldots ,\mathcal{K}_r$ that each contains $\frac{K}{r}\in \mathbb{Z}^+$ nodes. 
We define $\mathcal{T}\subset \mathcal K$ as a node group of size $r$ if it contains exactly one node from each $\mathcal{K}_i$, i.e., $|\mathcal{T}\cap\mathcal{K}_i|=1,\text{ }\forall \text{ } i\in [r]$. 
There are a total of $X=\left( \frac{K}{r} \right)^r$ possible node groups,  denoted by $\mathcal{T}_1,\ldots ,\mathcal{T}_{X}$. Furthermore, for each node group $\mathcal T_j$,  we define its $i$-th component $\mathcal T_{j,i}=\mathcal T_j \cap \mathcal K_i$ 
as the node in $\mathcal T_j$ that is chosen from $\mathcal K_i$, where $ i \in [r]$.



{\bf Node Group (NG) File Mapping}: Given node groups $\mathcal{T}_1,\ldots ,\mathcal{T}_{X}$, we split the $N$ files  into $X$ disjoint sets labeled as $\mathcal{B}_{1},\ldots,\mathcal{B}_{X}$. These file sets are of size $\eta_1\in \mathbb{Z}^+$ and $N=\eta_1 X$. Each file set $\mathcal{B}_{i}$ is only available  to every node in the node group  $\mathcal{T}_i$.
  It follows that if  node $k \in [K]$ belongs to a node group $\mathcal T_i$, then the file set 
  $\mathcal B_i$ is available to this node. 
  Hence, by considering all possible node groups $\mathcal T_i$ that node $k$ belongs to,  its available files, denoted by $\mathcal{M}_k$, is expressed as 
\be
\mathcal{M}_k:=\bigcup\limits_{i : k\in \mathcal{T}_i}\mathcal{B}_i.
\ee

{\bf Function Assignment 1}: The $Q$ reduce functions are split into $K$ equal size, disjoint subsets labeled as $\mathcal{W}_1, \ldots , \mathcal{W}_K$. Each set contains $\eta_2\in \mathbb{Z}^+$ reduce functions where $Q = \eta_2 K$. For each $k \in [K]$, define $\mathcal{W}_k$ as the set of reduce functions assigned to node $k$.  

\begin{remark}
  By Node Grouping 1 and NG File Mapping, each node set $\mathcal{K}_i$ collectively maps the file library exactly once, and therefore, the file library is mapped $r$ times among all $K$ nodes. Note that, since each file belongs to a unique file set $\mathcal B_i$ and is mapped to a unique set of $r$ nodes (in the node group $\mathcal T_i$), we must have $\frac{1}{N}\sum_{k=1}^K|\Mc_k|= \frac{N r}{N}=r$. Moreover, $\eta_1\left( \frac{K}{r} \right)^{r-1}$ files are mapped to each node. Then, by Function Assignment 1, each node is assigned $\eta_2$ reduce functions and each reduce function is assigned to exactly $s=1$ node.
\end{remark}

The Map, Shuffle and Reduce phases are defined as follows:

{\bf Map Phase}: Each node $k\in [K]$ computes the set of IVs $\{v_{i,j} : i\in [Q], w_j \in \mathcal{M}_k \}$.

{\bf Node Group (NG) Shuffle Method}:  For every $\alpha\in [X]$, a coded message will be multicasted by each node $k\in \mathcal{T}_\alpha$ to serve independent requests of the rest $r-1$ nodes in $\mathcal{T}_\alpha$. Meanwhile, each node $k\in \mathcal{T}_\alpha$ will multicast the same number of coded messages. 
Here, each IV is requested by a node $z \in \mathcal{T}_\alpha \setminus k$ and must be available to all other nodes in $\mathcal{T}_\alpha \setminus z$ to ensure that each node can decode successfully its own desired IVs from the broadcast. Next, we consider an arbitrary node group $\mathcal T_\alpha$ and a node $z \in \mathcal T_\alpha $. Assume that {$z \in \mathcal K_h$}, and thus  {$z=\mathcal T_{\alpha,h}= \mathcal T_\alpha \cap \mathcal K_h$}.
In the following, we fix the choice of $\alpha, z,h$ and
   define 
 \be
 \mathcal L_{z,\alpha}=\{ \ell \in [X]: \mathcal T_{\ell,h}\ne z, \mathcal T_{\ell,i}=\mathcal T_{\alpha,i},    \forall i\in[r]\setminus h \}.
 \label{eq:define_L_z_alpha}
 \ee
  Here, the set  $ \mathcal L_{z,\alpha}$  includes all indexes $\ell \in [X] $ such that the
  node group
  $\mathcal T_\ell$  differs from $\mathcal T_{\alpha}$ only in the $h$-th element, i.e., the node choice from {$\mathcal K_h$}. In other words, since {$z \in \mathcal K_h$}, then $\mathcal T_{\ell,h}$ can be any node in $\mathcal K_h$ except for $z$. Note that $h$ is suppressed from the subscript of $\mathcal L_{z,\alpha}$ for notation simplicity. The definition of (\ref{eq:define_L_z_alpha}) ensures that for any $\ell \in  \mathcal L_{z,\alpha}$, we have $z \notin \mathcal T_\ell$, but for any other node $z'$ in $\mathcal{T}_\alpha \setminus z$, we have $z' \in \mathcal T_\ell$. This follows that while file set $\mathcal B_\ell$ is not mapped to node $z$, it is mapped to all other nodes $z'$ in $\mathcal T_\alpha \setminus z$.
  Thus, we see that IVs of the type $\{v_{i,j}, i\in \mathcal{W}_z, w_j \in \mathcal B_\ell\}$ are requested by node $z$ because $z$ does not have $\mathcal B_\ell$, 
  but are available to all nodes in $\mathcal{T}_\alpha \setminus z$ because they all have access to  $\mathcal B_\ell$. 
   This  key idea is used to create multicast opportunities as follows.  
Formally, let us define
  \be
    \mathcal{V}_{\mathcal{T}_\alpha\setminus z}^{\{z\}}=\bigcup\limits_{\ell \in \mathcal L_{z,\alpha} }\left\{ v_{i,j}: i\in \mathcal{W}_z, w_j \in \mathcal B_\ell\right\},
    \label{eq:def_L_shuffle_1}
    \ee
 which contains IVs requested by node $z$ and are available at all nodes in $\mathcal{T}_\alpha\setminus z$.
      Furthermore, $\mathcal{V}_{\mathcal{T}_\alpha\setminus z}^{\{z\}}$ is split into $r-1$ disjoint subsets of equal size\footnote{In general, $|\mathcal{V}_{\mathcal{T}_\alpha\setminus z}^{\{z\}}|$ may not be divisible by $r-1$, in which case the IVs of $\mathcal{V}_{\mathcal{T}_\alpha\setminus z}^{\{z\}}$ can be concatenated into a message and split into $r-1$ equal size segments. This process was presented in \cite{li2018fundamental}.} denoted by $ \mathcal{V}_{\mathcal{T}_\alpha\setminus z}^{\{z\},\sigma_1},\ldots , \mathcal{V}_{\mathcal{T}_\alpha\setminus z}^{\{z\},\sigma_{r-1}} $ where $\{ \sigma_1,\ldots , \sigma_{r-1}\}=\mathcal{T}_\alpha\setminus z$. Each node $k\in \mathcal{T}_\alpha$ sends the common multicast message
    \be
    \label{eq: 1_trans_eq1}
    \bigoplus \limits_{z\in \mathcal{T}_\alpha\setminus k} \mathcal{V}_{\mathcal{T}_\alpha\setminus z}^{\{z\},k}
    \ee
    to all nodes  $z\in \mathcal{T}_\alpha\setminus k$.

{\bf Reduce Phase}: For all $k\in [K]$, node $k$ computes all output values $u_q$ such that $q\in \mathcal{W}_k$.

\begin{remark}
 For the homogeneous case, 
 we have $|\mathcal L_{z,\alpha}|=\frac{K}{r}-1$  because {$\mathcal T_{\ell,h}$} can only be one of the $\frac{K}{r}-1$ nodes in {$\mathcal K_h\setminus z$}.
When using Node Grouping 1, NG File Mapping, and Function Assignment 1, in NG Shuffle Method we find each intermediate value set, $\mathcal{V}_{\mathcal{T}_\alpha\setminus z}^{\{z\}}$, contains $\eta_1 \eta_2 \left( \frac{K}{r} - 1\right)$ IVs. 
\end{remark}


In the following, we will present a more complex $3$-dimension example by accommodating the design procedures and all the notations introduced above. 

\subsection{3-Dimension Example}
\label{sec: 3D_1}

\begin{figure}
\centering
\centering \includegraphics[width=9cm, height=6cm]{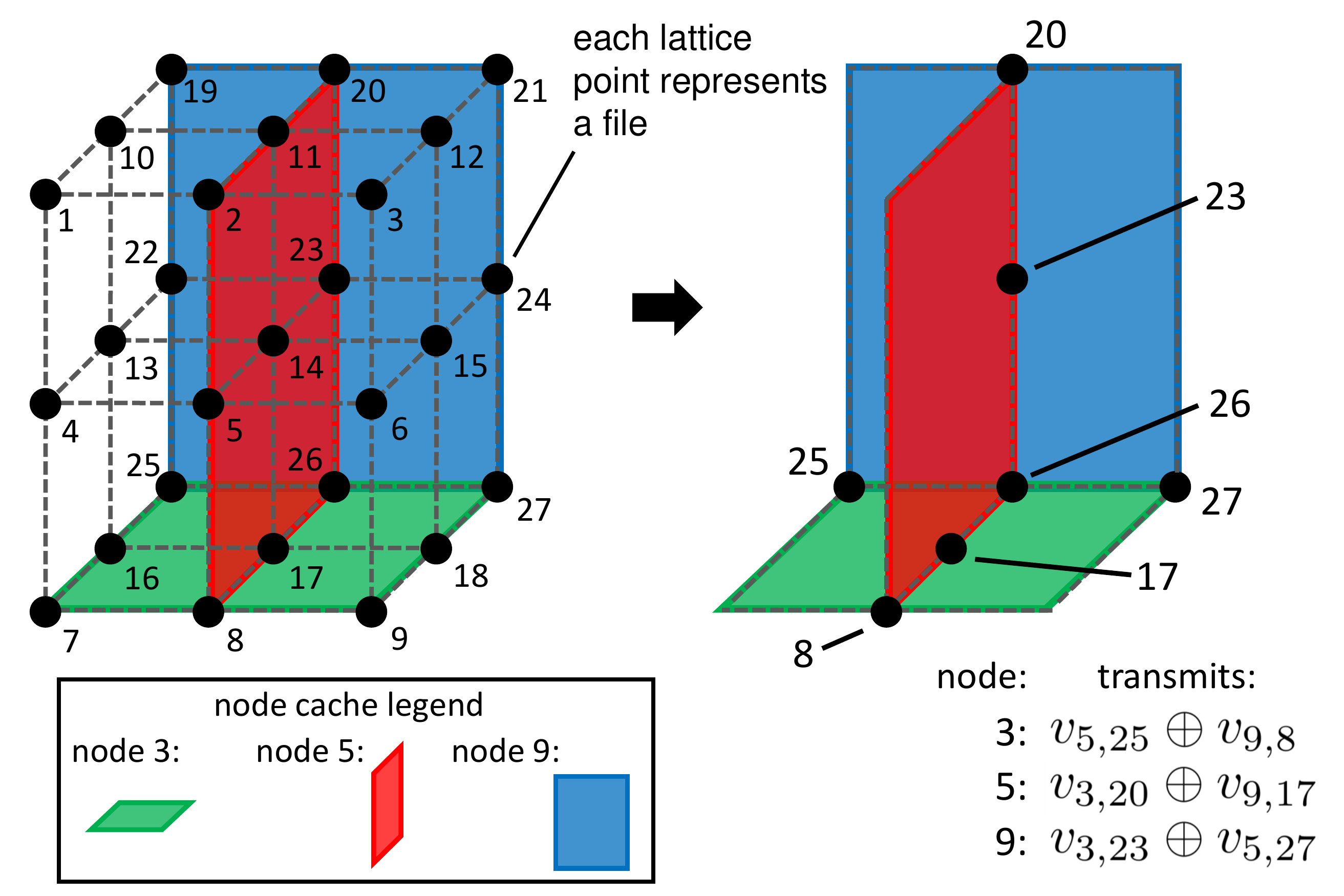}
\vspace{-0.6cm}
\caption{~\small  (left) Cube lattice which defines file availability amongst the $K=9$ computing nodes. Each lattice point represents a file and each node has set of files available to it represented by a plane of lattice points. The green, red and blue planes represent the files locally available to nodes $3$, $5$ and $9$, respectively. (right) Intersections of planes which represent files that are locally available to multiple nodes and yields coded multicasting opportunities.}
\label{fig: 3d fig}
\vspace{-0.4cm}
\end{figure}

To demonstrate the general scheme, we construct a computing network using a three-dimensional hypercube as shown in Fig.~\ref{fig: 3d fig}. 
Each lattice point in the cube, with its index $i\in [27]$ labeled next to the point, represents a different file set $\mathcal B_i=\{w_i\}$ which contains $\eta_1=1$ files. There are a total of  $K=9$ nodes, split into three node sets: $\mathcal{K}_1 = \{1,2,3 \}$,  $\mathcal{K}_2 = \{4,5,6 \}$, and $\mathcal{K}_3 = \{7,8,9 \}$, aligned along each of the $r=3$ dimensions of the hypercube. Specifically, the three nodes in $\mathcal{K}_1 = \{1,2,3 \}$ are represented by three parallel planes that go from top surface of the hypercube to the bottom.  Node 3 is represented by the green plane that passes through lattice point 7. Node 1 and 2 are represented by the two planes (not shown) parallel to the green plane that go through lattice point 1 and point 4, respectively. The three nodes in $\mathcal{K}_2 = \{4,5,6 \}$ are represented by
 three parallel planes that go from left surface of the hypercube to the right. Node 5 is represented by the middle plane, shown in red, that goes through lattice point 8,  and nodes 4 and 6 are represented by two planes (not shown) parallel to the red plane that go through lattice point 7 and 9, respectively.   The nodes in  $\mathcal{K}_3 = \{7,8,9 \}$ are represented by three parallel planes that go from the front surface of the hypercube to the back. Node 9 is the blue plane, passing through lattice point 27,  and nodes 7, 8 are represented by two planes (not shown) parallel to the blue plane and go through lattice points 9 and 18, respectively.   For file mapping, each node is assigned all the files indicated by the 9 lattice points on the corresponding plane. For instance,
  node $5$, represented by the red plane,  is assigned the file set $\mathcal{M}_5=\{ w_2, w_5, w_8, w_{11}, w_{14}, w_{17}, w_{20}, w_{23}, w_{26}\}$.  For each $i \in [3]$,  the size of $\mathcal K_i$ is $\frac{K}{r}=3$, which is the number of lattice points in the $i$-th dimension. Since the three nodes in each set  $\mathcal{K}_i$ are aligned along dimension $i$, they collectively store the entire  library of 27 files. Since each point $i$ in the lattice is uniquely determined by the intersection of three planes, one from each dimension, the same point also represents a node group $\mathcal T_i$. For instance, node group $\mathcal T_{26}=\{3,5,9\}$ is represented by the three planes-- green (node 3), red (node 5), and blue (node 9) intersecting at only one lattice point $i=26$. It is clear that each file $w_i$ is mapped to $r=3$ nodes in $\mathcal T_i$. Each node $k$ is assigned the $\eta_2=1$ functions of $\mathcal{W}_k=\{ k\}$ because $Q=K=9$ and each node $k$ is only assigned the $k$-th reduce function.



In the Map phase, each node computes all IVs from locally available files. 
 For example, node 5 will compute all possible IVs $\{ v_{i,j} : i \in[Q],w_j\in \mathcal{M}_5 \}$. The subset of IVs $\{ v_{5,j} : w_j\in \mathcal{M}_5 \}$ is used to calculate of the function output $u_5$. Furthermore, node $5$ will use the subset of IVs in $\{ v_{i,j} : i\in [Q]\setminus\mathcal{K}_2, w_j\in \mathcal{M}_5 \}$ for transmission and decoding purposes when forming multicasting groups with nodes of $\mathcal{K}_1$ and $\mathcal{K}_3$. Note that, similar to the last example, node $5$, and the other nodes, will compute some IVs that are not utilized.


We use the example of node group $\mathcal{T}_\alpha=\mathcal{T}_{26}=\{3,5,9\}$ to explain the Shuffle phase.
 Within node group $\mathcal{T}_{26}$, node 3 will multicast the summation of two IVs to nodes in $\mathcal T_{26}\setminus3$, one intended for node 5, and one intended for node 9. The former must be available at both nodes 3 and 9, and the latter must be available at both nodes 3 and 5. To determine these IVs, we consider the set $\mathcal{V}_{\{3,9\}}^{\{5\}} $ and $\mathcal{V}_{\{3,5\}}^{\{9\}}$. The set $\mathcal{V}_{\{3,9\}}^{\{5\}}$ contains two IVs requested by node 5 that are available at nodes $3,9$. To find these two IVs, we replace node $5$ in $\mathcal{T}_{26}$ by one of the other two nodes in $\mathcal K_2$, which are nodes 4 and 6. This way, we obtain two node sets $\mathcal T_{25}=\{3,4,9\}$ and  $\mathcal T_{27}=\{3,6,9\}$  that differ from $\mathcal{T}_\alpha$ only in the second element (the element that intersects $\mathcal K_2$).
  Thus ${\mathcal L_{5,\alpha}}=\{25, 27\}$.  This leads to $\mathcal{V}_{\{3,9\}}^{\{5\}}  =\mathcal{V}_{\{3,9\}}^{\{5\},3} \bigcup \mathcal{V}_{\{3,9\}}^{\{5\},9} =\{ v_{5,25},v_{5,27} \}$ which contains two IVs requested by node $5$ and are available at nodes 3 and 9. Similarly, we find ${\mathcal L_{9,\alpha}}=\{8,17\}$ and
$\mathcal{V}_{\{3,5\}}^{\{9\}} =\mathcal{V}_{\{3,5\}}^{\{9\},3} \bigcup\mathcal{V}_{\{3,5\}}^{\{9\},5}=\{ v_{9,8},v_{9,17} \} $. Once these two IV sets are found, node 3 transmits the summation of one IV from each set, say $\mathcal{V}_{\{3,9\}}^{\{5\},3} \oplus \mathcal{V}_{\{3,5\}}^{\{9\},3}= v_{5,25} \oplus v_{9,8}$ to nodes 5 and 9. Upon receiving this value, node 5 will subtract $v_{9,8}$ to recover $v_{5,25}$ and node 9 will subtract $v_{5,25}$ to recover $v_{9,8}$. The rest of the IV sets  can be found in a similar fashion such that   $\mathcal{V}_{\{5,9\}}^{\{3\}} = \mathcal{V}_{\{5,9\}}^{\{3\},5}  \bigcup  \mathcal{V}_{\{5,9\}}^{\{3\},9} =\{ v_{3,20},v_{3,23} \}$,  and $\mathcal{V}_{\{3,5\}}^{\{9\}}= \mathcal{V}_{\{3,5\}}^{\{9\},3} \bigcup \mathcal{V}_{\{3,5\}}^{\{9\},5}  = \{ v_{9,8}, v_{9,17} \}$. Node 5 will transmit $v_{3,20} \oplus v_{9,17}$ to nodes 3 and 9. Node 9 will transmit $v_{3,23} \oplus v_{5,27}$ to nodes 3 and 5.


In this example,
each node participates in $9$ multicasting groups and transmits $1$ coded message per group. Each transmission has the equivalent size of $1$ IV. Therefore, the communication load is $L_{\rm c}=\frac{9 \cdot 9}{QN}= \frac{81}{9\cdot 27}=\frac{1}{3}$, which is half of the uncoded communication load $L_{\rm u}=\frac{2}{3}$, or the fraction of files not available to each node.

\subsection{Achievable trade-off between Computation and Communication Loads}


The following theorem evaluates the trade-off between the computation and communication loads for the proposed scheme.

\begin{theorem}
\label{theorem: s1_hom}
By using Node Grouping 1, NG File Mapping, Function Assignment 1, and NG Shuffle Method, the communication load of the general homogeneous 
scheme is
\begin{eqnarray}
\label{eq: theorem s1_hom}
L_{\rm c}(r) &=& 
\frac{K-r}{K\left( r-1 \right)}.
\end{eqnarray}
\hfill $\square$
\end{theorem}
\begin{IEEEproof}
Theorem~\ref{theorem: s1_hom} is proved in Appendix \ref{sec: codedHetPrfs1} as a special case of our general heterogeneous design which is defined in Section \ref{sec: hets1}. 
\end{IEEEproof}

The optimality of this scheme is discussed in Section~\ref{sec: Discussion} by comparing the communication load of this scheme with that of the state-of-the-art scheme in \cite{li2018fundamental}.

\section{Heterogeneous Hypercube Computing Approach}
\label{sec: hets1}
In this section, we expand the proposed combinatorial hypercube design to accommodate heterogeneous computing networks. As mentioned in the introduction, one key novelty of our design is nodes are assigned a varying number of files and reduce functions so that, in practice, nodes with more computational capability perform relatively more of the overall MapReduce execution. In this case, 
the proposed heterogeneous design becomes a hypercuboid, consisting of $P$ interleaved homogeneous hypercube networks. The homogeneous networks, $\mathcal{C}_p, \; \forall p\in[P]$, reflect hypercubes with different dimensions and lengths, representing distinct classes of nodes with varying storage capacity and computation resources. 
We start with an example and then present the general scheme.

\subsection{3-Dimension Hypercuboid Example}

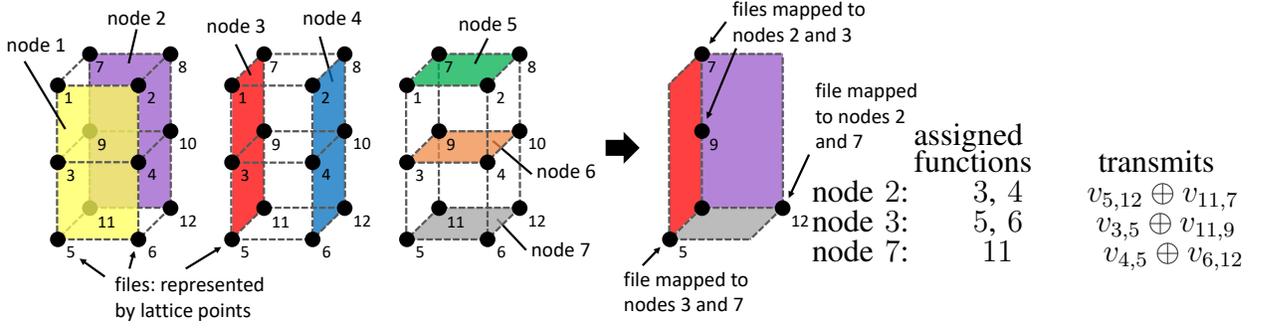
\begin{figure*}
\centering
\centering \includegraphics[width=16cm]{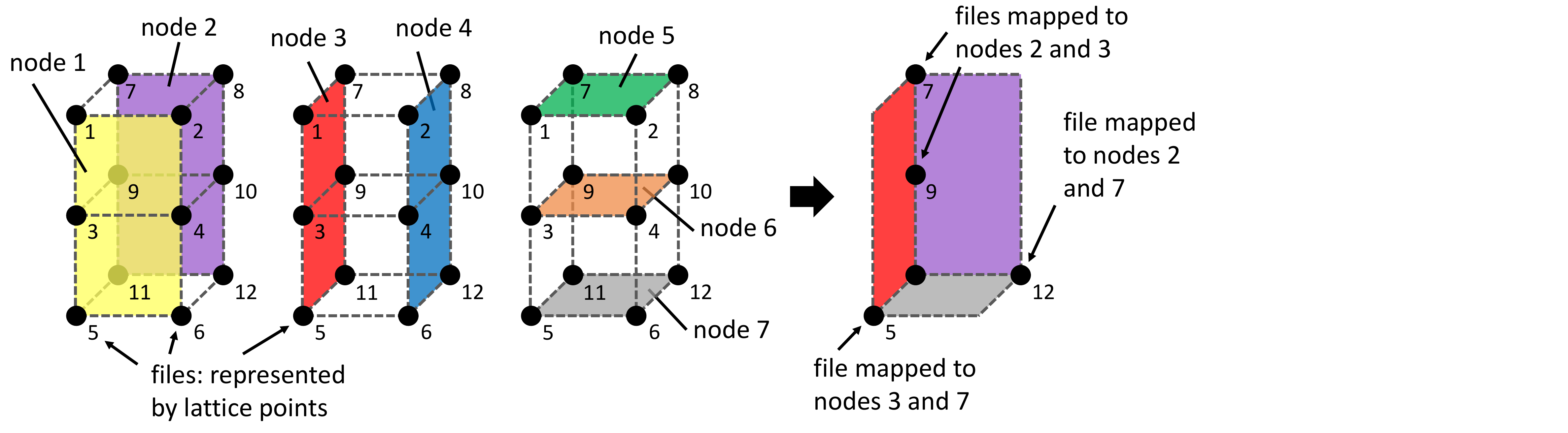} 
\put(-38.4,24.5){assigned}
\put(-38.4,21){functions \;\;\;\;\; transmits}
\put(-51.95,17){node $2$: \;\;\;\;\;\;$3$, $4$ \;\;\;\;\;\;$v_{5,12}\oplus v_{11,7}$ }
\put(-51.95,13){node $3$: \;\;\;\;\;\;$5$, $6$ \;\;\;\;\;\;\;$v_{3,5}\oplus v_{11,9}$}
\put(-51.95,9){node $7$: \;\;\;\;\;\;\;$11$ \;\;\;\;\;\;\;\;\;$v_{4,5}\oplus v_{6,12}$ }
\vspace{-0.2cm}
\caption{~\small Representations of a hypercuboid with $P=2$, $r_1=2$, $m_1=2$, $r_2=1$ and $m_2=3$.  {\it Left $3$ cuboids}: Depictions of the files mapped to the nodes of $\mathcal{K}_1$, $\mathcal{K}_2$ and $\mathcal{K}_3$, respectively. {\it Right most cuboid}: Hypercuboid highlighting the files stored at exactly $2$ nodes of the multicast group $\mathcal{T}_{11}=\{ 2,3,7 \}$. These files, in addition to the assigned functions among these nodes, determine the IVs included in the coded multicasts, which are displayed to the right.}
\label{fig: het_s1_exp1}
\vspace{-0.4cm}
\end{figure*}

This example is presented in Fig.~\ref{fig: het_s1_exp1}, where 
there are two classes of nodes $\mathcal C_1=\mathcal K_1 \cup \mathcal K_2$ and $\mathcal C_2=\mathcal K_3$ with different storage capability where $\mathcal{K}_1~=~\{ 1, 2 \}$, $\mathcal{K}_2 = \{ 3, 4 \}$ and $\mathcal{K}_3 = \{ 5, 6, 7 \}$. Each node in $\mathcal C_1$  stores half of the files and each node in $\mathcal C_2$  stores one-third of the files. Each node set, $\mathcal{K}_i$, collectively  stores all $N=12$ files. Each file is assigned to a node group $\mathcal{T}_\alpha$ of $3$ nodes such that it contains one node from each set $\mathcal{K}_1$, $\mathcal{K}_2 $ and $\mathcal{K}_3$. For example, file $w_1$ is assigned to the nodes of $\{1,3,5 \}$ and file $w_{11}$ is assigned to the nodes of $\{2,3,7 \}$. All of the files assignments are represented by the cuboid in Fig.~\ref{fig: het_s1_exp1}. In the Map phase, the nodes will compute all IVs from their locally available files. Since every file is assigned to $3$ nodes, the computation load is $r=3$.

Different from previous works in CDC, nodes are assigned a varying number of reduce functions. We assign more reduce functions to nodes which have larger storage and computing capability.  Assume that there are $Q=11$ reduce functions. We assign $2$ reduce functions to each node of $\mathcal{K}_1 $ and $\mathcal{K}_2$ and just $1$ reduce function to each node of $\mathcal{K}_3$. Specifically, the function assignments are $\mathcal W_1=\{1,2\}$, $\mathcal W_2=\{3,4\}, \mathcal W_3=\{5,6\}, \mathcal W_4=\{7,8\}$, and $\mathcal W_5=\{9\}, \mathcal W_6=\{10\}$, and $\mathcal W_7=\{11\} $.
    The reason we assigned this specific number of reduce functions to each node will become clear when we discuss the Shuffle phase.

In the Shuffle phase, the set of multicast groups includes all possible node groups $\mathcal T_{\alpha}$ which contain 1 node from each set $\mathcal{K}_1$, $\mathcal{K}_2 $ and $\mathcal{K}_3$.
Within each $\mathcal T_{\alpha}$, nodes send coded pairs of IVs to the other two nodes. For example, consider the node set $\mathcal T_{\alpha}=\mathcal T_{11}=\{2,3,7 \}$. Following notations in Shuffle Method 1, we have $\mathcal L_{2,\alpha}=\{5\}$. This is because when replacing node $2 \in \mathcal K_1$ in $\mathcal T_{\alpha}$ by a different node in  $\mathcal K_1$, we obtain $\mathcal T_{5}=\{1,3,7 \}$. Hence, using $\mathcal W_2=\{3,4\}$ and Eqn. (\ref{eq:def_L_shuffle_1}), we obtain $\mathcal V^{\{2\}}_{3,7}=\{v_{3,5}, v_{4,5}\}$, which are IVs requested by node 2 and computed at nodes $3$ and $7$. Similarly, for node 3, we have $\mathcal L_{3,\alpha}=\{12\}$, and $\mathcal V^{\{3\}}_{2,7}=\{v_{5,12}, v_{6,12}\}$. For node 7, we have $\mathcal L_{7,\alpha}=\{7,9\}$, corresponding to $\mathcal T_{7}=\{2,3,5\}$ and $\mathcal T_{9}=\{2,3,6\}$. While the size of $\mathcal L_{7,\alpha}$ is larger than that of $\mathcal L_{2,\alpha}$ and $\mathcal L_{3,\alpha}$, since $\mathcal W_7=\{11\}$ is smaller, we obtain $\mathcal V^{\{7\}}_{2,3}=\{v_{11,7}, v_{11,9}\}$, which is the same size as that of $\mathcal V^{\{2\}}_{3,7}$ and $\mathcal V^{\{3\}}_{2,7}$. Using Eqn.(\ref{eq: 1_trans_eq1}), we see that nodes $2$, $3$, and $7$  transmit $v_{5,12}\oplus v_{11,7}$,  $v_{3,5}\oplus v_{11 ,9}$, and $v_{4,5}\oplus v_{6,12}$, respectively.


In this example, we see that by assigning a varying number of reduce functions to the nodes we can  create symmetry among each node group, $\mathcal{T}_\alpha$, i.e., each node of the group requests the same number of IVs from the other nodes of the group. This symmetry can lead to savings in the communication load.
Here, the communication load can be calculated by accounting for the $2\cdot 2 \cdot 3 = 12$ node groups, where within each group, there are $3$ transmissions of size $T$ bits. By normalizing by $QNT$ we find the communication load of the coded scheme is $L_{\rm c} = \frac{36}{12 \cdot 11} = \frac{3}{11}$. We can compare this to the uncoded communication load, where each requested IV is transmitted alone. To compute the uncoded communication load, we count the number of IVs each node requests. Since the $4$ nodes of $\mathcal{K}_1 $ and $\mathcal{K}_2$ request $6\cdot 2 = 12$ IVs each and the $3$ nodes of $\mathcal{K}_3$ request $8$ IVs each, we find $L_{\rm u} = \frac{4\cdot 12 + 3 \cdot 8}{12\cdot 11} = \frac{6}{11}$. In this case, $L_{\rm c} = \frac{1}{2}\cdot L_{\rm u}$ since for the coded Shuffle policy every requested IV is transmitted in coded pairs. In the general heterogeneous CDC scheme proposed here, we will see that $L_c = \frac{1}{r-1}\cdot L_{\rm u}$.


\subsection{General Heterogeneous Scheme} 
\label{sec: gen_het s1}

In this subsection, we will introduce the general heterogeneous scheme for 
step by step. 

{\bf Node Grouping 2}: The key idea of Node Grouping 2 is to form one heterogeneous network based on a hypercuboid design that consists of  $P$ interleaved homogeneous networks, represented by hypercubes of different dimensions $r_p$ and sizes $m_p$ within the hypercuboid.    
{The $K$ nodes consist of $P$ disjoint sets denoted by $\mathcal{C}_1,\ldots ,\mathcal{C}_P$, where $\sum_{p=1}^{P}|\mathcal{C}_p|=K$.
 For each $p\in [P]$, split $\mathcal{C}_p$ into  $r_p \in \mathbb{Z}^+$ disjoint subsets, each of size $m_p$, denoted by 
 $\{\mathcal{K}_{n_{p\scaleto{-1}{3.5pt}}+1}
 ,\ldots , \mathcal{K}_{n_p}\}$, where $n_p=\sum_{i=1}^p r_i$.
  Hence, the entire network is comprised of $r$ node sets, $\mathcal{K}_1 ,\ldots , \mathcal{K}_r$, where $r = \sum_{p=1}^{P}r_p$.  
  Consider all possible node groups $\mathcal{T}_1,\ldots ,\mathcal{T}_{X}$ of size $r$ that each  contains one node from every node set $\mathcal{K}_1 ,\ldots , \mathcal{K}_r$, here  $X=\prod_{i=1}^{r}|\mathcal{K}_i|=\prod_{p=1}^{P}m_p^{r_p}$.  Denote $\mathcal T_{j,i}=\mathcal T_j \cap \mathcal K_i, \;\forall j \in [X]$ and $\forall i \in [r]$, as the node in $\mathcal T_j$ that is chosen from $\mathcal K_i$.
}

The file mapping is then determined by the NG  File Mapping  defined in Section \ref{sec: genscheme_s1} with node groups
$\mathcal{T}_1,\ldots ,\mathcal{T}_{X}$ defined by Node Grouping 2.


\begin{remark}
  When using Node Grouping 2 and NG File Mapping, we form a hypercuboid made of $P$ interleved hypercubes of different dimensions. For a given $p\in [P]$, $\mathcal{C}_p$ translates to $r_p$ dimensions of size $m_p$ of the hypercuboid. Moreover, $\mathcal{C}_p$ serves the role that is similar to that of a single hypercube of dimension $r_p$ as in the homogeneous case. Specifically, $\mathcal{C}_p$
   contains $r_p$ node sets $\mathcal K_i$, each of size $m_p$. Here $m_p$ is the number of lattice points along each dimension of  the hypercube.
   The total number of nodes in $\mathcal C_p$ is thus $r_p m_p$.
   Nodes in each $\mathcal K_i$  collectively map the file library once. Hence, all nodes in $\mathcal{C}_p$  have the same storage capacity that each maps a total of $\frac{N}{m_p}$ files.
   Collectively, nodes in $\mathcal{C}_p$ map the library $r_p$ times.  The $P$ disjoint sets of $\mathcal C_1, \cdots, \mathcal C_p$ form one hypercuboid with $r$ dimensions where there are $r_p$ dimensions of size $m_p$ for $p\in[P]$.
 Hence, each node group $\mathcal{T}_\alpha$ of size $r=\sum_{i=p}^P r_p$, defined in Node Group 2, consists of  the union of $P$ node groups, 
with size $r_1, \cdots, r_P$, respectively, chosen  from each of the $P$ interleved hypercubes corresponding to $\mathcal C_p, p \in[P]$. Note that, instead of each hypercube operating independently subject to its own computation load, $r_p$, the hypercuboid design takes full advantage of the total computation load, $r$, across the $P$ hypercubes to achieve the gain of $\frac{1}{r-1}$ for the heterogeneous system.
\end{remark}

{\bf Function Assignment 2}:
Define $Y$ as the least common multiple (LCM) of $\{m_1-1, m_2-1, \ldots , m_P-1 \}$. Split the $Q$ functions into $K$ disjoint sets, labeled 
$\mathcal{W}_{1},\ldots,\mathcal{W}_{K}$,
where, in general, the sets may be different sizes. For each $k \in [K]$,  $|\mathcal{W}_{k}| =~\frac{\eta_2Y }{m_p - 1}$ where $k \in \mathcal{C}_p$ and $\eta_2 \in \mathbb{Z}^+$ such that $Q~=~\eta_2 Y \sum_{p=1}^{P}\frac{r_p m_p}{m_p - 1}$. For each $k \in [K]$, 
let $\mathcal{W}_k$ be
 the set of reduce functions assigned to node $k$.

The Map and Shuffle phases 
follow our standard definition from Section \ref{sec: genscheme_s1} and the NG Shuffle Method is used for the Shuffle phase with node grouping defined by Node Grouping 2.

The correctness of the proposed heterogeneous CDC scheme is proved in Appendix \ref{sec_APP:correctness}. 

\begin{remark}
When using Node Grouping 2, NG File Mapping,  Function Assignment 2, and  NG Shuffle Method,  we find that each intermediate value set $\mathcal{V}_{\mathcal{T}_\alpha\setminus z}^{\{z\}}$ contains $\eta_1 \eta_2 Y$ IVs.
\end{remark}

\begin{remark}
Node Grouping 2 and Function Assignment 2 are a more general case of Node Grouping 1 and Function Assignment 1, respectively. Therefore, the homogeneous scheme of Section \ref{sec: genscheme_s1} is a special case of the general heterogeneous scheme here. By letting $P=1$ such that $\mathcal{C}_1$ is the set of all nodes, we find $r=r_1$, $m_1 = \frac{K}{r}$, $X = \left( \frac{K}{r} \right)^r$ and $Y = \frac{K}{r}-1$. Moreover, each node is assigned $\frac{\eta_2 Y}{m_1-1} = \eta_2$ reduce functions. For file availability, nodes are split into $r$ disjoint, equal size sets, $\mathcal{K}_1,\ldots,\mathcal{K}_r$, and file sets of size $\eta_1$ are available to sets of nodes which contain exactly one node from each set $\mathcal{K}_1,\ldots,\mathcal{K}_r$.
\end{remark}

\begin{remark}
It can be seen that the proposed hypercuboid design may not work for any given heterogeneous individual memories and computation loads due to the constrained combinatorial structure. In practice, we can group  nodes with heterogeneous storage capacity and computation resources to fit a hypercuboid design as close as possible (similar to ``quantization") to reap the benefit by taking the heterogeneity of the system into the consideration.
\end{remark}

%
%

\subsection{Achievable Trade-off between Computation and Communication Loads}

In this section, we first present the communication load of an uncoded Shuffle phase, $L_{\rm u}$, 
using Node Grouping 2, NG File Mapping, Function Assignment 2 of the general heterogeneous scheme.
Here, uncoded Shuffle phase means that all the requested IVs will be transmitted in a unicast fashion without coded multicasting. Note that, $L_{\rm u}$ represents the fraction of intermediate values which are requested by any node. Then, we demonstrate that the communication load using the  
the proposed hypercuboid scheme 
and the NG Shuffle Method
is $L_{\rm c} = \frac{1}{r-1}\cdot L_{\rm u}$. More formally, we define $L_{\rm u}$ and $L_{\rm c}$ as functions of $m_1 , \ldots , m_P$ and $r_1 , \ldots , r_P$ which define the number of nodes and the corresponding computation load in each node class of the heterogeneous computing network. Then, $L_u$ and $L_c$ are given in the following theorems. 

\begin{theorem}
\label{theorem: uncoded}
By using Node Grouping 2, NG File Mapping, Function Assignment 2, and an uncoded Shuffle phase, the communication load is
\begin{align}
\label{eq: Lu}
L_{\rm u}(m_1 , \ldots , m_P, r_1 , \ldots , r_P) = \frac{r}{\sum_{p=1}^{P}\frac{r_p m_p}{m_p-1}}.
\end{align}
\end{theorem}

%

\begin{IEEEproof}
Theorem \ref{theorem: uncoded} is proven in Appendix \ref{sec: uncodedHetPrfs1}. 
\end{IEEEproof}
The following theorem states the communication load of the Shuffle phase which uses coded communication.

\begin{theorem}
\label{theorem: coded}
By using Node Grouping 2, NG File Mapping, Function Assignment 2, and NG Shuffle Method, the communication load of the general heterogeneous 
scheme is
\begin{align}
\label{eq: Lc}
L_{\rm c} (m_1 , \ldots , m_P, & r_1 , \ldots , r_P) = \frac{1}{r-1}\cdot\frac{r}{\sum_{p=1}^{P}\frac{r_p m_p}{m_p-1}} = \frac{1}{r-1} \cdot L_{\rm u}(m_1 , \ldots , m_P, r_1 , \ldots , r_P).
\end{align}
\end{theorem}

\begin{IEEEproof}
Theorem \ref{theorem: coded} is proven in Appendix \ref{sec: codedHetPrfs1}. 
\end{IEEEproof}
The communication load $L_{\rm c}$ is comprised of two parts: the local computing gain, $L_{\rm u}$, and the global computing gain, $\frac{1}{r-1}$. The local computing gain represents the normalized number of IVs that must be shuffled. As nodes have access to a larger fraction of the files, the nodes will inherently request less in the Shuffle phase. The global computing gain stems from the fact that with the coded design every transmission serves $r-1$ nodes with distinct requests.

\subsection{Optimality}
The information theoretic lower bound of the communication load derived in \cite{li2018fundamental} is under the assumption of the homogeneous reduce function assignment. Hence, it 
does not apply when reduce functions are heterogeneously assigned to the computing nodes. In the following we discuss the lower bound of the communication load for two scenarios. First, we demonstrate a straightforward lower bound on communication load when considering all possible file and function assignments for a given $r$ and $K$. Next, we provide a lower bound on the communication load when we use the specific file and function assignments (Node Grouping 2, NG File Mapping and Function Assignment 2) of the heterogeneous design in Section \ref{sec: gen_het s1}.

A trivial bound on the communication load is $L \geq 0$. Given $r$ and $K$, the following file and function assignment and Shuffle phase design will yield a communication load meeting this bound. Pick $r$ nodes and assign the entire file library to each of the nodes. Furthermore, for each function, assign it to one of the $r$ nodes with access to the entire file library. As every node that is assigned a reduce function is able to compute all the necessary IVs itself, no Shuffle phase is required such that $L=0$. Note that, in this context, we do not consider any storage or computing limitations on the nodes, rather, we show that optimizing the communication load over all possible function and file assignments is not an interesting problem. 

The question remains as to the optimality of the proposed Shuffle phase of Section \ref{sec: gen_het s1} given the file and reduce function assignments. Based on the seminal approach introduced in \cite{arbabjolfaei2013capacity,wan2016caching,wan2016optimality,wan2020index} for coded caching with uncoded cache placement, we derive Theorem \ref{theorem: bound} which provides a lower bound on the entropy of all transmissions in the Shuffle phase given a specific function and file placement and a permutation of the computing nodes. 

\begin{theorem}
\label{theorem: bound}
Given a particular file placement, $\mathcal{M}_k,\;\forall k\in[K]$ and function assignment $\mathcal{W}_k,\;\forall k\in[K]$, in order for every node $k\in[K]$ to have access to all IVs necessary to compute functions of $\mathcal{W}_k$, the optimal communication load over all achievable shuffle schemes, $L^*$, is bounded by
  \be \label{eq: bound_eq1}
  L^* \geq  \frac{1}{TQN}\sum_{i=1}^{K} H\left(V_{\mathcal{W}_{k_i},:}|V_{:,\mathcal{M}_{k_i}},Y_{\{k_1,\ldots, k_{i-1} \}}\right)
  \ee
  where $k_1, \ldots , k_K$ is some permutation of $[K]$, $V_{\mathcal{W}_{k_i},:}$ 
  is the set of IVs necessary to compute the functions of $\mathcal{W}_{k_i}$,\footnote{The notation ``$:$" is used to denote all possible indices.} $V_{:,\mathcal{M}_{k_i}}$ is set of IVs which can be computed from the file set $\mathcal{M}_{k_i}$ and $Y_{\{k_1,\ldots, k_{i-1} \}}$  is the union of the set of IVs necessary to compute the functions of  $\bigcup_{j=1}^{i-1}\mathcal{W}_{k_j}$ and the set of IVs which can be computed from files of $\bigcup_{j=1}^{i-1}\mathcal{M}_{k_j}$.
\hfill $\square$
\end{theorem}
\begin{IEEEproof}
Theorem \ref{theorem: bound} is proved in Appendix \ref{appendix: bound}. 
\end{IEEEproof}

In  Theorem \ref{thm: optimality} below, we demonstrate that
given Node Grouping 2, NG File Mapping, and Function Assignment 2, 
the NG  Shuffle Method 
introduced in Section~\ref{sec: genscheme_s1} 
yields a communication load that is within a constant of the lower bound.

\begin{theorem}
\label{thm: optimality}
For a computing network defined by 
Node Grouping 2, NG File Mapping, and Function Assignment 2, 
define $L^*$ to be the infimum of the communication load over all possible Shuffle phases, then we have
\be
 L_{\rm c} \leq \frac{2r}{r-1} L^*,
\ee
where $L_{\rm c}$, given in (\ref{eq: Lc}), is the communication load achieved by using the NG Shuffle Method. 
\end{theorem}
\begin{IEEEproof}
Theorem \ref{thm: optimality} is proved in  Appendix \ref{sec: opt_pf}. 
\end{IEEEproof}

\section{Discussions}
\label{sec: Discussion}
In this section, we will compare the performance the proposed schemes to the state-of-the-art schemes in \cite{li2018fundamental}. Specifically, we compare the required number of files, the required number of multicast groups 
and the communication load. 
When we compare the performance of the proposed heterogeneous CDC scheme with that of the homogeneous CDC in \cite{li2018fundamental}, we fix the computation load, $r$, the number of files, $N$, and the number of reduce functions, $Q$.\footnote{We adjust $N$ and $Q$ to be the same by using the appropriate $\eta_1$ and $\eta_2$.} 

The scheme in \cite{li2018fundamental} requires $N_1 = {K \choose r}\eta_1$ input files, $Q_1 = {K \choose s}\eta_2$ reduce functions. Moreover, the communication load as a function of $K$, $r$ and $s$ is \be \label{eq: li_se1} 
L_1(r) = \frac{1}{r}\left(1-\frac{r}{K}\right). 
\ee

\subsection{Homogeneous CDC} 

\begin{figure}
\vspace{-5cm}
\centering
\centering \includegraphics[width=12cm]{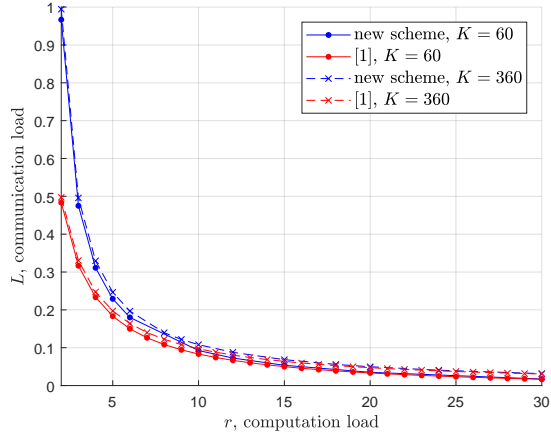}
\vspace{-5cm}
\caption{~\small A comparison of the resulting communication load for the newly proposed and the state-of-the-art homogeneousdistributed computing schemes.}
\label{fig: s1graph}
\vspace{-0.5cm}
\end{figure}

Using (\ref{eq: theorem s1_hom}), we observe the following comparison
\be
\frac{L_{\rm c}(r)}{L_1(r)} = \frac{rK}{K-r} \cdot\frac{K-r}{K\left( r-1 \right)} = \frac{r}{r-1}.
\ee
For most values of $r$ there is an insignificant increase in the communication load for the new combinatorial scheme and furthermore for $r \rightarrow \infty$ the two schemes yield the identical communication loads. 
Since our proposed homogeneous scheme uses the same function assignment as the scheme in \cite{li2018fundamental}, then 
this hypercube based design is asymptotically optimal in the information theoretic sense in general without fixing the file and function assignments. These findings are verified through simulation of the communication load as shown in Fig.~\ref{fig: s1graph}.

While both schemes require the same number of outputs functions, $Q=K\eta_2$, the required number of input files has been drastically reduced in this case. 
It can be observed that the number of input files for the homogeneous hypercube design is 
\be
N_{\rm c} =  \left( \frac{K}{r} \right)^r\eta _1
\ee
while the scheme of \cite{li2018fundamental} requires $N_1 = {K \choose r}\eta_1$ input files. Assuming $r  =\Theta (K)$, by use of Stirling's formula to directly compare the two equations yields
\begin{align} 
\frac{N_1}{N_{\rm c}} &= \frac{{K \choose r}}{\left( \frac{K}{r}\right)^r} =  \frac{K!}{r!(K-r)!\left( \frac{K}{r}\right)^r} = \Theta\left( \frac{\sqrt{2\pi K}\left( \frac{K}{e}\right)^K}{2 \pi \sqrt{r \left( K-r \right)}\left( \frac{r}{e}\right)^r\left( \frac{K-r}{e}\right)^{\left( K-r \right)}} \cdot \frac{1}{\left( \frac{K}{r}\right)^r} \right)  \notag\\
&= \Theta\left(\sqrt{\frac{K}{2\pi r (K-r)}}\cdot \left( \frac{K}{K-r} \right)^K \right) = \Theta\left(\sqrt{\frac{1}{K}}\cdot \left( \frac{K}{K-r} \right)^K \right)\label{eq: comp_s1_files}.
\end{align}
When $r<K$, we find that (\ref{eq: comp_s1_files}) grows exponentially with $K$ and, therefore, our proposed scheme has an exponential decrease in the number of required files.

As pointed out in \cite{li2018fundamental,konstantinidis2018leveraging}, the required number of multicast group is also an important design parameter in CDC. If this number is large, it may take a long time to build such node groups such that the gain achieved by CDC is completely gone. It can be seen that the number of required multicast groups for the scheme in \cite{li2018fundamental} is $U_1={K \choose r+1}$, while the required number of multicast group of the proposed scheme is $U_c = \left(\frac{K}{r}\right)^r$. Hence, by a similar computation to (\ref{eq: comp_s1_files}), it can be seen that 
\be
\frac{U_1}{U_c} = \Theta\left(\frac{r+1}{K-r} \cdot \sqrt{\frac{1}{K}}\cdot \left( \frac{K}{K-r} \right)^K \right),
\ee
which can grows exponentially with $K$ such that the proposed hypercube scheme reduces the required number of multicast group exponentially. 

\begin{remark}
The hypercube approach has similar performance compared to the CDC scheme based on the resolvable design proposed in \cite{konstantinidis2018leveraging}, e.g., the required number of input files in  \cite{konstantinidis2018leveraging} is $\left(\frac{K}{r}\right)^{r-1}$, which is slightly better than the proposed hypercube scheme. However, as we discussed in Section~\ref{sec: hets1}, the proposed hypercube scheme can be extended to the heterogeneous CDC networks naturally while it is unclear how to extend the scheme in \cite{konstantinidis2018leveraging} to heterogeneous CDC networks. 
\end{remark}


\subsection{Heterogeneous CDC} 
\label{sec: disc_sg1}


As shown in (\ref{eq: Lc}), the communication load of the proposed heterogeneous CDC design is $L_c(r)=\frac{1}{r-1}L_{\rm u}(r)$, where $\frac{1}{r-1}$ and $L_{\rm u}(r)$  are  the global computing gain and the local computing gain, respectively.   In comparison, for the homogeneous design in \cite{li2018fundamental}, we have $L_1(r)=\frac{1}{r} (1-\frac{r}{K})$, where the global computing gain is $\frac{1}{r}$ and the local computing gain is $1-\frac{r}{K}$. Next, we will show that even though the proposed heterogeneous design has an inferior global computing gain than that of  \cite{li2018fundamental} ($\frac{1}{r-1}$ versus $\frac{1}{r}$), it has a better local computing gain $L_{\rm u}(r)\le (1-\frac{r}{L})$, and hence can have a better communication load $L_c(r) < L_1(r)$ under certain parameter regimes. 

Since $\sum_{p=1}^{P} \frac{r_p}{r} = 1$ and $\frac{m_p}{m_p-1}$ is a convex function of $m_p$ for $m_p > 1$, using (\ref{eq: Lu}) and Jensen's inequality, we can obtain 
\begin{align}
\frac{1}{L_{\rm u}(r)} = \frac{\sum_{p=1}^{P}\frac{r_p m_p}{m_p-1}}{r} &= \sum_{p=1}^{P} \frac{r_p}{r}\cdot \frac{m_p}{m_p-1}
\geq \frac{\sum_{p=1}^{P} \frac{r_p m_p}{r}}{\left(\sum_{p=1}^{P} \frac{r_p m_p}{r}\right) - 1}
= \frac{\frac{K}{r}}{\frac{K}{r}-1}
= \frac{K}{K-r}
\label{eq:Lu}
\end{align}
where $\sum_{p=1}^{P}r_p m_p = \sum_{p=1}^{P}|\mathcal{C}_p| = K$. 
{Note that the inequality in (\ref{eq:Lu}) is strictly ``$>$'' if the network is truly heterogeneous, i.e., not all $\{m_p\}$ are equal.}
Hence, 
\be
L_{\rm u}(r) \leq \frac{K-r}{K} = 1-\frac{r}{K}, 
\label{eq:upper_bound_local_gain}
\ee
which shows that the local computing gain for our heterogeneous design is upper bounded by that  of the homogeneous design in \cite{li2018fundamental}.
Using  (\ref{eq: Lc}), we obtain, 
\be
L_{\rm c}(r)=\frac{1}{r-1}L_{\rm u}(r) \leq \frac{1}{r-1}\cdot \left( 1-\frac{r}{K} \right).
\ee
 Thus, $L_{\rm c}(r)$ can be less than $L_1(r)$ for certain choices of $r$ and $K$. For example, given a heterogeneous network defined by $m_1 = 2$, $r_1 = 4$ and $m_2 = 8$, $r_2 = 2$, we have $r=r_1+r_2=6$, $K=r_1m_1+r_2m_2=24$. 
 We compare it with a homogeneous network with $r=6$ and $K=24$.  The proposed heterogeneous design has a local computing gain of $L_{\rm u}(r) = \frac{7}{12} \approx 0.583$, which is less than that of the homogeneous design $1-\frac{r}{K} = \frac{3}{4} = 0.75$, and a communication load of $L_{\rm c} = \frac{7}{ 60}\approx 0.117$, that is lower than that of the homogeneous design
 $L_1(r) = \frac{1}{8} = 0.125$.

\begin{remark}
  In \cite{li2018fundamental}, $L_1(r)$ was proved to be a lower bound on the communication load given $r$ and $K$. However, the proof uses the implicit assumption that every node is assigned the same number of reduce functions. Our new finding is that if the reduce functions can be assigned in a heterogeneous fashion, then the communication load lower bound of \cite{li2018fundamental} does not apply.
\end{remark}

\begin{figure}
\vspace{-5cm}
\centering
  \includegraphics[width=12cm]{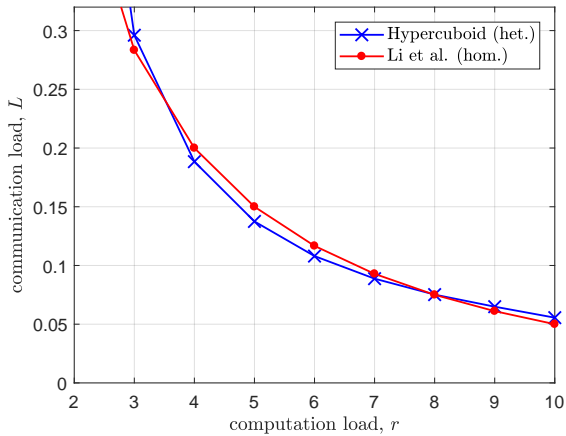}
  \vspace{-5cm}
\caption{\small A comparison of the proposed hypercuboid CDC design to the state-of-the-art CDC design of \cite{li2018fundamental} with $K=20$ nodes and an equivalent computation load, $r$. The heterogeneous hypercube is designed with parameters $r_1=r-1$, $m_1=2$, $r_2=1$ and $m_2=K-2(r-1)$. The hypercuboid design has a lower communication load than that of the homogeneous design for for $4\leq r \leq 7$.}
\label{fig: het sim_seq1_fig}
\vspace{-0.5cm}
\end{figure}

In Fig.~\ref{fig: het sim_seq1_fig}, we provide additional comparisons of the communication load of the hypercuboid design and  the homogeneous scheme of \cite{li2018fundamental} with an equivalent computation load, $r$. Each design has a fixed number of nodes $K=20$. The heterogeneous network is defined with $P=2$ sets of nodes that map a different number of files and are assigned a different number of reduce functions. Specifically, there are $|\mathcal{C}_1|=2(r-1)$ powerful nodes and $|\mathcal{C}_2|=K-2(r-1)$ weaker nodes where $r_1=r-1$, $m_1=2$, $r_2=1$ and $m_2=K-2(r-1)$. In other words, the nodes of $\mathcal{C}_1$ each map $\frac{1}{2}$ of the files and the nodes of $\mathcal{C}_2$ each map a $\frac{1}{K-2(r-1)}$ fraction of the files which can be much less than $\frac{1}{2}$. Fig.~\ref{fig: het sim_seq1_fig} shows that the communication load of the hypercuboid design is less than that of the state-of-the-art homogeneous design of \cite{li2018fundamental} for $4\leq r \leq 7$.

{\bf Comparisons for large networks.} 
Next, we provide comparisons of the communication load of  the proposed heterogeneous scheme and the homogeneous scheme \cite{li2018fundamental} for networks with a large number of computing nodes $K$. We consider two cases. 

Case 1. For the heterogeneous network, assume that $r_1,\ldots , r_P$ and $r$  are fixed, but  the fraction of files each node has access to, $\frac{1}{m_1}, \cdots, \frac{1}{m_P}$, decrease as $K$ becomes large. Then, we have
\be
\lim_{K\rightarrow \infty} L_{\rm u}(r)=1 \quad \text{and} \quad
\lim_{K\rightarrow \infty} \frac{L_{\rm c}(r)}{L_1(r)} = \frac{r}{r-1}.
\ee
In other words, $\frac{L_{\rm c}(r)}{L_1(r)} = \Theta (1)$.

Case 2. 
For the heterogeneous network, assume that
$\frac{r_1}{K}=\beta_1,\ldots , \frac{r_P}{K}=\beta_K$ and $\frac{r}{K}=\beta$ are kept constant as $K$ gets large. The fraction of files  available to each node,  $\frac{1}{m_1}, \cdots, \frac{1}{m_P}$,  are also kept constant.
It then follows from (\ref{eq:Lu}) that when the network is truly heterogeneous (not all $\{m_p\}$ are equal), then we have
\be
\lim_{K\rightarrow \infty} \frac{L_{\rm c}(r)}{L_1(r)} = \lim_{r\rightarrow \infty} \frac{r}{r-1} \cdot \frac{L_{\rm u}(r)}{1-\frac{r}{K}} =\frac{1}{1-\beta} \Bigg(\frac{1}{  \sum_{p=1}^P \frac{\beta_p}{\beta} \frac{m_p}{m_p-1} }\Bigg) < \frac{1}{1-\beta} (1-\beta)= 1.
\ee
This means that for large networks considered here, the communication  load of the proposed heterogeneous scheme is {\it strictly less} than that of the homogeneous scheme. Hence, for some heterogeneous file and computation load assignments, the fundamental trade-off proposed in \cite{li2018fundamental} is ``breakable".  As we discussed before, in the extreme case, where there exists a ``super node" that can store all the files and compute all functions, the communication load is straightforwardly $0$. However, for given heterogeneous storage capacities and computation loads, it is non-trivial to design an achievable CDC scheme such that its performance  is superior compared to that of homogenous CDC under the same total storage and computation load constraint. 

For the hypercuboid design, the required number of files is $N=X=\prod_{p=1}^{P}m_p^{r_p}$ and reduce functions is $Q~=~ Y \sum_{p=1}^{P}\frac{r_p m_p}{m_p - 1}$ where $Y$ is the LCM of $\{m_1-1,\ldots ,m_P-1 \}$. 
Unlike the homogeneous network case, due to the lack of CDC design for general heterogeneous networks, we cannot compare the proposed scheme to other schemes. Nevertheless, we believe that these numbers can serve as a benchmark for the future research in this topic. 

\section{Conclusions and Future Directions}
\label{sec: Conclusion}


In this work, we introduced a novel hypercuboid combinatorial approach to design CDC for both homogeneous and heterogeneous distributed computing networks.  
This new design achieves a significant reduction in the number of files and functions compared to the state-of-the-art scheme in \cite{li2018fundamental}. Moreover, the proposed schemes maintain a multiplicative computation-communication trade-off and are proven to be asymptotically optimal. {Most importantly, 
we provided
an explicit and systematic heterogeneous CDC design with optimality guarantees
under certain network parameters.}
Surprisingly, {we found} that the optimal trade-off derived in \cite{li2018fundamental} no longer applies when functions are heterogeneously assigned and {as a result}, the communication load of a heterogeneous network can be less than that of an equivalent homogeneous CDC network. 
For the future research direction, first, it will be interesting to design other achievable schemes with heterogeneous function assignments and a more general communication load bound given a set of storage capacity requirements of  computing nodes. Second, it is challenging but important to characterize the information theoretic converse given the storage capacity and the computation load constraints of each node without fixing the file and output function assignments. 
\appendices

\section{Proof of Theorems \ref{theorem: s1_hom} and \ref{theorem: coded}}
\label{sec: codedHetPrfs1}
\begin{comment}
For any $n \in [X]$, and for all $z \in \mathcal{T}_n$ where $z \in \mathcal{K}_p$, we find
\begin{align}
\big| & \mathcal{V}_{\mathcal{T}_n\setminus z}^{\{z\}} \big| = \left | \mathcal{W}_z \right | \times \left | \left \{ w_j: w_j \notin \mathcal{M}_z, w_j \in \bigcap\limits_{k \in \mathcal{T}_n\setminus z} \mathcal{M}_k , \right \} \right| \\
&= \left | \mathcal{W}_z \right | \cdot \eta_1 \left | \left \{ \mathcal{T}_{n'} : \{\mathcal{T}_n\setminus z \} \subset \mathcal{T}_{n'}, z \notin \mathcal{T}_{n'}, n' \in [X] \right \} \right | \\
&= \left | \mathcal{W}_z \right | \cdot \eta_1 \left | \left \{ \mathcal{T}_{n'} : \{\mathcal{T}_n\setminus z \} \cup k = \mathcal{T}_{n'}, k \in \mathcal{K}_p\setminus z,  \right \} \right | \\
&= \left | \mathcal{W}_z \right | \cdot \eta_1 \left | \left \{ k :  k \in \mathcal{K}_p\setminus z,  \right \} \right | \\
&= \left | \mathcal{W}_z \right | \cdot \eta_1 (|\mathcal{K}_p|-1) = \frac{\eta_2Y }{m_p - 1} \cdot \eta_1 (m_p - 1)= \eta_1 \eta_2 Y.
\end{align}
\end{comment}
For any $\alpha \in [X]$, and  $z \in \mathcal{T}_\alpha$, where $z \in \mathcal{K}_h \subseteq \mathcal{C}_p$, it follows from  Eq. (\ref{eq:define_L_z_alpha}), (\ref{eq:def_L_shuffle_1}),   and Remark 3 in Section \ref{sec: genscheme_s1} that 
\begin{equation}
\big|  \mathcal{V}_{\mathcal{T}_n\setminus z}^{\{z\}} \big| = \left | \mathcal{W}_z \right | \cdot \eta_1 \big| \mathcal L_{z,\alpha} \big|=\left | \mathcal{W}_z \right | \cdot \eta_1 (|\mathcal{K}_p|-1) = \frac{\eta_2Y }{m_p - 1} \cdot \eta_1 (m_p - 1)= \eta_1 \eta_2 Y .
\end{equation}
We consider $X$ node groups of size $r$ nodes, where for each group, every node of that group transmits a coded message of size $\big|  \mathcal{V}_{\mathcal{T}_n\setminus z}^{\{z\}} \big| / (r-1)$, therefore, the communication load is
\begin{align}
L_{\rm c} ( m_1 , \ldots & , m_P, r_1 , \ldots , r_P) = \frac{1}{QN}\cdot X \cdot r \cdot  \frac{\big|  \mathcal{V}_{\mathcal{T}_n\setminus z}^{\{z\}} \big|}{r-1} \\
&= \frac{1}{\left(\eta_2 Y \sum_{p=1}^{P}\frac{r_p m_p}{m_p - 1}\right)\eta_1 X}\cdot X \cdot r \cdot \frac{\eta_1 \eta_2 Y}{r-1} = \frac{1}{r-1}\cdot\frac{r}{\sum_{p=1}^{P}\frac{r_p m_p}{m_p-1}}.
\end{align}

For the special homogeneous case, where $P=1$ and $\mathcal{C}_1$ is the set of all nodes, we find $r=r_1$, $m_1 = \frac{K}{r}$ and
\begin{align}
L_{\rm c} &= \frac{1}{r-1}\cdot\frac{r}{\sum_{p=1}^{P}\frac{r_p m_p}{m_p-1}} =\frac{1}{r-1}\cdot\frac{r}{\left(\frac{K}{\frac{K}{r}-1}\right)} =\frac{1}{r-1}\cdot\frac{K-r}{K}.
\end{align}

Hence, we finished the proof of Theorems \ref{theorem: s1_hom} and \ref{theorem: coded}. 

\section{Proof of Theorem \ref{theorem: uncoded}}
\label{sec: uncodedHetPrfs1}
For all $p \in [P]$,  the number of files a node $k~\in~\mathcal{K}_j~\subseteq~\mathcal{C}_p$ has local access to is
\begin{align}
|\mathcal{M}_k| &= \eta_1 \prod\limits_{i\in[r]\setminus j}|\mathcal{K}_i| = \frac{\eta_1 X}{|\mathcal{K}_j|} = \frac{N}{m_p}.
\end{align}
We count the number of IVs that are requested by any node and normalize by $QN$
\begin{align}
L_u & (m_1 , \ldots , m_P, r_1 , \ldots , r_P) = \frac{1}{QN}\sum_{k\in[K]}  | \left \{ v_{i,j} : i \in \mathcal{W}_k , w_j \notin \mathcal{M}_k  \right \} | \\
&= \frac{1}{QN}\sum_{k\in[K]} \left | \mathcal{W}_k \right|\times \left( N- \left |  \mathcal{M}_k   \right| \right) = \frac{1}{QN}\sum_{p\in[P]} \sum_{k\in\mathcal{C}_p} \left | \mathcal{W}_k \right|\times \left( N- \left |  \mathcal{M}_k   \right| \right) \\
&= \frac{1}{QN}\sum_{p\in[P]} \sum_{k\in\mathcal{C}_p} \frac{\eta_2Y }{m_p - 1} \cdot \left( N- \frac{N} {m_p} \right) = \frac{1}{Q}\sum_{p\in[P]} r_p m_p \frac{\eta_2Y }{m_p - 1}\left( \frac{m_p-1}{m_p} \right) \\
&= \frac{\eta_2 Y \sum_{p\in[P]} r_p }{\eta_2 Y \sum_{p=1}^{P}\frac{r_p m_p}{m_p - 1} } = \frac{r}{\sum_{p=1}^{P}\frac{r_p m_p}{m_p-1}}
\end{align}
where $|\mathcal{C}_p| = r_p m_p$ for all $p \in [P]$. 
Hence, we finished the proof of Theorem  \ref{theorem: uncoded}. 

\section{Correctness of Heterogeneous 
CDC Scheme}
\label{sec_APP:correctness}
This proof includes 4 parts: 1) nodes only compute IVs from locally available files, 2) nodes only transmit locally computed IVs, 3) nodes can decode transmissions with requested IVs and 4) after the Map and Shuffle phases, nodes have all necessary IVs to compute their reduce functions.

For 1), any node $k\in [K]$ computes intermediate values of the set
\be
\label{eq: pf4}
\{ v_{i,j} : i \in [Q] , w_j \in \mathcal{M}_k \}
\ee
In all cases $w_j \in \mathcal{M}_k$ for any $v_{i,j}$ computed by node $k$, therefore nodes only compute IVs from locally available files.
\begin{comment}
We demonstrate 2) and 3) simultaneously by observing a node group, $\mathcal{T}_n$ for any $n \in [X]$, in the Shuffle phase. A node $k \in \mathcal{T}_n$ has computed the intermediate values of $\mathcal{V}_{\mathcal{T}_n\setminus z}^{\{z\}}$ for all $z\in \mathcal{T}_n\setminus k$. This is true because
\be
 \bigcap \limits_{a\in\mathcal{T}_n\setminus z}\mathcal{M}_a  \subseteq  \mathcal{M}_k
\ee
where $k \in \mathcal{T}_n \setminus z$ and, therefore,
\begin{multline}\label{eq: pf3}
\mathcal{V}_{\mathcal{T}_n\setminus z}^{\{z\}}=\left\{ v_{i,j}: i\in \mathcal{W}_z,w_j \notin \mathcal{M}_z , \text{ }w_j\in \bigcap \limits_{a\in\mathcal{T}_n\setminus z}\mathcal{M}_a \right\}  \subseteq
\{ v_{i,j} : i \in [Q] , w_j \in \mathcal{M}_k  \}.
\end{multline}
The the right hand side of (\ref{eq: pf3}) matches that of (\ref{eq: pf4}) and defines a set of intermediate values that node $k$ computes in the Map phase. Since $\mathcal{V}_{\mathcal{T}_n\setminus z}^{\{z\}}$ is computed by node $k$ for all $z\in \mathcal{T}_n\setminus k$, it is clear that node $k$ can transmit coded messages consisting of subsets of $\mathcal{V}_{\mathcal{T}_n\setminus z}^{\{z\}}$  for all $z\in \mathcal{T}_n\setminus k$ and this satisfies 2).
\end{comment}

Next, we prove 2) and 3) simultaneously. Consider any node group $\mathcal{T}_\alpha$ and any node $k \in \mathcal{T}_\alpha$. We need to confirm that node $k$ has access to the multicast messages defined in Eq. (\ref{eq:def_L_shuffle_1}) and (\ref{eq: 1_trans_eq1}). This is true because as discussed above Eq. (\ref{eq:def_L_shuffle_1}), all nodes in $\mathcal{T}_\alpha \setminus z$, including node $k$,  have access to the file set $\mathcal B_\ell$ where $\{\mathcal{T}_\alpha \setminus z\} \subset \mathcal{T}_\ell$. To see 3), when a node $z_0 \in \mathcal T_\alpha$ receives a multicast message from another node $k \in \mathcal T_\alpha$ that takes the form of Eq. (\ref{eq:def_L_shuffle_1}), only one term, $\mathcal{V}_{\mathcal{T}_\alpha\setminus z}^{\{z_0\},k}$, is its desired message. The other terms are of the form $\mathcal{V}_{\mathcal{T}_\alpha\setminus z}^{\{z\},k}$, intended for node $z$, where $z\in \mathcal T_\alpha$ and , $z\ne z_0,k$. Since node $z_0 \in \mathcal T_\alpha \setminus z$, it has access to  $\mathcal{V}_{\mathcal{T}_\alpha\setminus z}^{\{z\},k}$, and thus can decode its desired message correctly. 
\begin{comment}
In Appendix \ref{sec: codedHetPrfs1} we show that $\left |\mathcal{V}_{\mathcal{T}_n\setminus z}^{\{z\}}\right | = \eta_1 \eta_2 Y$ for any $z\in [K]$ and $n$ such that $z \in \mathcal{T}_n$. Furthermore, node $k$ receives transmissions with requested intermediate values from all nodes $z\in \mathcal{T}_n\setminus k$ where each transmission consists of $r-1$ coded intermediate value sets which are the same size. Only one of these sets are a subset of $\mathcal{V}_{\mathcal{T}_n \setminus k}^{\{ k\}}$ and the rest are subsets of $\mathcal{V}_{\mathcal{T}_n\setminus z}^{\{z\}}$ for some $z\in \mathcal{T}_n\setminus k$. Therefore, node $k$ can recover its requested intermediate value set since the other intermediate value sets are locally computed and this satisfies 3).
\end{comment}

To prove 4), we need to show that for a given $z\in \mathcal K_h$, if some file $w_j \notin \mathcal M_z$, then node $z$ must be able to recover its desired IVs $\{v_{i,j}: i\in \mathcal W_z\}$ from multicast messages of the form Eq. (\ref{eq:def_L_shuffle_1}) and (\ref{eq: 1_trans_eq1}). 
To see this, assume that $w_j \in \mathcal B_{\ell_0}$. Consider node group $\mathcal T_{\ell_0}$. Since $w_j \in \mathcal B_{\ell_0}$ and $w_j \notin \mathcal M_z$, we must have $z \notin \mathcal T_{\ell_0}$. In other words, $\mathcal T_{\ell_0,h} \ne z$. Now, consider another node group $z \in \mathcal T_\alpha$ such that $\mathcal T_\alpha$ and $\mathcal T_{\ell_0}$ differs only in the $h$-th element: $\mathcal T_{\alpha,h}=z$ and $\mathcal T_{\alpha,i}=\mathcal T_{\ell_0,i}$ for any $i \ne h$. As defined in Eq. (\ref{eq:def_L_shuffle_1}),  since $\ell_0 \in \mathcal L_{z,\alpha}$ and $w_j \in \mathcal B_{\ell_0}$, node $z$ will be able to received its desired IVs  $\{v_{i,j}: i\in \mathcal W_z\}$ from the multicast group messages from node group $\mathcal T_\alpha$ according to  Eq. (\ref{eq:def_L_shuffle_1}) and (\ref{eq: 1_trans_eq1}).

\section{Proof of Theorem \ref{theorem: bound}}
\label{appendix: bound}
In this proof, we use the following notation: $\mathcal{K}$ is the set of all nodes, $X_{\mathcal{K}}$ represents the collection of all transmissions by all nodes in $\mathcal{K}$, $\mathcal{W}_{\mathcal{S}}$ is the set of functions assigned to at least one node of $\mathcal{S}$, $\mathcal{M}_{\mathcal{S}}$ is the set files locally available to at least one node in $\mathcal{S}$, $V_{\mathcal{W}_{\mathcal{S}_1},\mathcal{M}_{\mathcal{S}_2}}$ is the set of IVs needed to compute the functions of $\mathcal{W}_{\mathcal{S}_1}$ and computed from the files of $\mathcal{M}_{\mathcal{S}_2}$. Finally, we define the following
\be
Y_\mathcal{S} \triangleq \left(V_{\mathcal{W}_\mathcal{S},:},V_{:,\mathcal{M}_\mathcal{S}}\right)
\ee
where ``$:$" is used to denote all possible indices.

%
Given all the transmissions from all nodes, $X_\mathcal{K}$, and IVs which can be locally computed by a node $k$, $V_{:,\mathcal{M}_k}$, node $k$ needs to have access to all IVs necessary for its assigned functions, $V_{\mathcal{W}_k,:}$, therefore
\be
H(V_{\mathcal{W}_k,:} | X_\mathcal{K}, V_{:,\mathcal{M}_k}) = 0.
\ee

Given this assumption, we find
\begin{align}
H(X_\mathcal{K}) &\geq H(X_\mathcal{K}|V_{:,M_{k_1}}) 
= H(X_\mathcal{K},V_{\mathcal{W}_{k_1},:}|V_{:,M_{k_1}}) - H(V_{\mathcal{W}_{k_1},:} | X_\mathcal{K}, V_{:,\mathcal{M}_{k_1}}) \nonumber\\
&= H(X_\mathcal{K},V_{\mathcal{W}_{k_1},:}|V_{:,M_{k_1}}) \nonumber\\
&= H(V_{\mathcal{W}_{k_1},:}|V_{:,M_{k_1}}) + H(X_\mathcal{K}|V_{\mathcal{W}_{k_1},:},V_{:,M_{k_1}})\nonumber\\
&=H(V_{\mathcal{W}_{k_1},:}|V_{:,M_{k_1}}) + H(X_\mathcal{K}|Y_{k_1}). \label{eq: claim 1 2}
\end{align}
Similarly,
\begin{align}
H&(X_\mathcal{K}|Y_{\{k_1,\ldots k_{i-1}\}}) 
\geq H(X_\mathcal{K}|V_{:,M_{k_i}},Y_{\{k_1,\ldots k_{i-1}\}}) \nonumber\\
&= H(X_\mathcal{K},V_{\mathcal{W}_{k_i},:}|V_{:,M_{k_i}},Y_{\{k_1,\ldots k_{i-1}\}}) 
- H(V_{\mathcal{W}_{k_i},:} | X_\mathcal{K}, V_{:,\mathcal{M}_{k_i}},Y_{\{k_1,\ldots k_{i-1}\}}) \nonumber 
\end{align}
\begin{align}
&= H(X_\mathcal{K},V_{\mathcal{W}_{k_i},:}|V_{:,M_{k_i}},Y_{\{k_1,\ldots k_{i-1}\}}) \nonumber\\ 
&= H(V_{\mathcal{W}_{k_i},:}|V_{:,M_{k_i}},Y_{\{k_1,\ldots k_{i-1}\}}) 
+ H(X_\mathcal{K}|V_{\mathcal{W}_{k_i},:},V_{:,M_{k_i}},Y_{\{k_1,\ldots k_{i-1}\}})\nonumber\\
&= H(V_{\mathcal{W}_{k_i},:}|V_{:,M_{k_i}},Y_{k_1,\ldots k_{i-1}}) + H(X_\mathcal{K}|Y_{\{k_1,\ldots k_i\}}). \label{eq: claim 1 3}
\end{align}
Also, since nodes can only transmit IVs from locally available files, we see that 
$H(X_\mathcal{K}|Y_{\{k_1,\ldots k_K\}}) = 0$.
By starting with (\ref{eq: claim 1 2}) and iteratively using the relationship of (\ref{eq: claim 1 3}) to account for all $k_i \in \mathcal{K}$, we obtain
\be \label{eq: th 4 ent sum}
H(X_\mathcal{K}) \geq \sum_{i=1}^{K} H\left(V_{\mathcal{W}_{k_i},:}|V_{:,\mathcal{M}_{k_i}},Y_{\{k_1,\ldots, k_{i-1} \}}\right).
\ee
Moreover, since $H(X_\mathcal{K}) = LTQN$, from (\ref{eq: th 4 ent sum}) we obtain the lower bound on the optimal communication load, $L^*$, of (\ref{eq: bound_eq1}) and proved Theorem \ref{theorem: bound}.

\section{Proof of Theorem \ref{thm: optimality}}
\label{sec: opt_pf}

We define a permutation of the $K$ nodes, $(k_1, \ldots , k_K)$, such that $\{ k_1 , \ldots , k_{m_p} \} = \mathcal{K}_i \subseteq \mathcal{C}_p$ for some $i\in[r]$ and $p\in[P]$ as defined in Section~\ref{sec: gen_het s1}. 
For $1 \leq j \leq m_p$, given all IVs collectively computed by nodes $k_1 , \ldots , k_j$ and all IVs needed by nodes $k_1 , \ldots , k_{j-1}$ to compute their respective reduce functions, the entropy of the requested IVs of the node $k_j$ is
\begin{align}
H &\left( \mathcal{V}_{ \mathcal{W}_{k_j} , :} | \mathcal{V}_{ : , \mathcal{M}_{k_1} } , Y_{ \{ k_1,\ldots k_{j-1} \} }\right) = H \left( \mathcal{V}_{ \mathcal{W}_{k_j} , :} | \mathcal{V}_{ : , \mathcal{M}_{\{ k_1 , \ldots , k_{j-1} \}} } \right) = T |\mathcal{W}_{k_j}|\left( N - \bigcup\limits_{j' \in [j]} \mathcal{M}_{k_{j'}} \right) \nonumber \\
&= T\cdot\frac{\eta_2Y}{m_p-1}\left( N - \sum_{j' \in [j]}|\mathcal{M}_{k_{j'}}| \right) =\frac{T\eta_2Y}{m_p-1}\left( N - \frac{jN}{m_p} \right) = \frac{T\eta_2YN}{(m_p-1)m_p}\left( m_p - j \right).
\end{align}
Furthermore, since the nodes $k_1 , \ldots , k_{m_p}$ collectively have access to all the $N$ files and compute all $QN$ intermediate values, we see that for $m_p \leq j \leq K$
\be
H \left( \mathcal{V}_{ \mathcal{W}_{k_j} , :} | \mathcal{V}_{ : , \mathcal{M}_{k_1} } , Y_{ \{ k_1,\ldots k_{j-1} \} }\right) = 0.
\ee
By using of the bound of Theorem \ref{theorem: bound}
\begin{align}
L^* &\geq \frac{1}{QNT}\sum_{j=1}^{m_p-1} H \left( \mathcal{V}_{ \mathcal{W}_{k_j} , :} | \mathcal{V}_{ : , \mathcal{M}_{k_1} } , Y_{ \{ k_1,\ldots k_{j-1} \} }\right) =\frac{1}{Q}\sum_{j=1}^{m_p-1}\frac{\eta_2Y}{(m_p-1)m_p}\left( m_p - j \right) \nonumber\\
&= \frac{\eta_2Y}{Q(m_p-1)m_p}\sum_{j=1}^{m_p-1} j = \frac{\eta_2Y}{Q(m_p-1)m_p}\cdot \frac{m_p(m_p-1)}{2} = \frac{\eta_2Y}{2Q}= \frac{1}{2\sum_{p=1}^{P}\frac{r_p m_p}{m_p-1}}.
\end{align}
Finally, we see that
\be
\frac{L_{\rm c}}{L^*} \le \frac{2r}{r-1} \leq 4
\ee
for $r\geq 2$. This completes the proof of Theorem \ref{thm: optimality}.

\bibliographystyle{IEEEbib}
\bibliography{references_d2d}

\end{document}